\documentclass[10pt]{article}
\usepackage{ulem}
\usepackage{times}
\usepackage{color}
\usepackage{graphicx}
\usepackage{calc}
\usepackage{epsfig}
\usepackage{float}
\usepackage{amsfonts}
\newfloat{Continuation}{htb}{lob}
 \fussy \oddsidemargin0.0cm \evensidemargin0.0cm
\begin{document}

\begin{center}
{\LARGE \bf Burst firing is a neural code in an insect auditory system.}
\end{center}
\vspace{0.5cm}

\begin{center}
{\bf  Hugo G. Eyherabide$^{1,2}$, Ariel Rokem$^{1}$, Andreas V. M.
Herz$^{1}$, and In\'es Samengo$^{2,^*}$}
\end{center}

\noindent $^1$ Institute for Theoretical Biology, Department of Biology, Humboldt Universit\"at,
and
Bernstein Center for \\
$\phantom{^1}$ Computational Neuroscience, 10115 Berlin, Germany.\\
$^2$ Centro At\'omico Bariloche and Instituto Balseiro, 8400 San Carlos de Bariloche, Argentina.

\vspace{1cm} \noindent {\bf Running title:}\\ Burst firing is a neural code. \vspace{1cm}

\noindent {\bf Correspondence:} \\In\'es Samengo\\ Centro At\'omico Bariloche\\
San Carlos de Bariloche, (8400), R\'{\i}o Negro, Argentina.\\ Tel:
++ 54 2944 445100 (int: 5391 / 5345). Fax: ++54 2944 445299.\\
Email: samengo@cab.cnea.gov.ar

\pagebreak

\oddsidemargin0.0cm \evensidemargin0.0cm

\section*{Abstract}

\noindent Various classes of neurons alternate between
high-frequency discharges and silent intervals. This phenomenon is
called burst firing. To analyze burst activity in an insect system,
grasshopper auditory receptor neurons were recorded {\em in vivo}
for several distinct stimulus types. The experimental data show that
both burst probability and burst characteristics are strongly
influenced by temporal modulations of the acoustic stimulus. The
tendency to burst, hence, is not only determined by cell-intrinsic
processes, but also by their interaction with the stimulus time
course. We study this interaction quantitatively and observe that
bursts containing a certain number of spikes occur shortly after
stimulus deflections of specific intensity and duration. Our
findings suggest a sparse neural code where information about the
stimulus is represented by the number of spikes per burst,
irrespective of the detailed interspike-interval structure within a
burst. This compact representation cannot be interpreted as a
firing-rate code. An information-theoretical analysis reveals that
the number of spikes per burst reliably conveys information about
the amplitude and duration of sound transients, whereas their time
of occurrence is reflected by the burst onset time. The investigated
neurons encode almost half of the total transmitted information in
burst activity.

\section*{Keywords}

\noindent Burst spiking, neural code, sensory encoding, information
theory, auditory receptor.

\pagebreak

\section{Introduction}
\label{s1}

\noindent Tonic and burst firing encode different
aspects of the sensory world. Specifically, in thalamic relay cells,
burst firing has been reported as more efficient in signal detection
than tonic firing (Grubb and Thompson, 2005; Lesica et al., 2006;
Sherman 2001) and more reliable to repeated presentations of the
same stimulus (Alitto et al., 2005; Denning and Reinagel, 2005).
Tonic firing, in turn, seems to be well suited for encoding the
detailed evolution of time-varying stimuli. Similar results have
been obtained in electric fish (Chacron et al., 2004; Metzner et
al., 1998; Oswald et al., 2004).

Various studies have compared the stimuli that trigger isolated
spikes with those that induce burst firing (Alitto et al., 2005;
Denning and Reinagel, 2005; Eggermont and Smith, 1996;  Grubb and
Thompson, 2005; Metzner et al., 1998; Oswald et al., 2004; Reinagel
et al., 1999). In these comparisons bursts were taken as a single
type of event, without further discrimination between different
burst variants. However, bursts may also encode stimuli in a graded
manner (Kepecs et al., 2001; Oswald et al., 2007; Kepecs et al.,
unpublished). Bursts with different numbers of spikes can thus act
as compact {\em code-words}. Indeed, in neurons from various sensory
systems the number $n$ of spikes within a burst correlates with
particular properties of the external stimulus, such as the
orientation of a drifting sine-wave grating (DeBusk et al., 1997)
and the slope or the amplitude of visual contrast changes (Kepecs et
al., 2001; Kepecs et al., unpublished).

Here, we examine the role of bursts in grasshopper auditory receptor
cells. When stimulated with time-dependent acoustic signals, these
neurons fire high-frequency bursts that are triggered by stimulus
deflections of specific intensity and duration. We quantify the
amount of information encoded by a burst code and characterize the
stimulus features represented by bursts of different duration.
Receptor cells, however, do not generate bursts in response to
constant or step stimuli (Gollisch et al., 2002; Gollisch and Herz,
2004), indicating that bursts can result from a nontrivial interplay
between external stimuli and intrinsic dynamics. Our analysis leads
to the following conclusions: (a) burst-firing constitutes a
prominent feature in the neural code of the investigated auditory
neurons, (b) representing neural responses by {\em intra-burst spike
counts} $n$ allows one to estimate the amount and type of
transmitted information in a straightforward manner, (c) the
correspondence between code-words and the stimulus features that
they represent may be readily explored with burst-triggered
averages. Most importantly, (d) burst coding is a key element in the
transmission of time-varying stimuli even for cells that are not
intrinsic bursters.


\section{Methods}
\label{s2}

\subsection{Electrophysiology and stimulus design}
\label{s2p1}

All experiments were conducted on adult {\em Locusta Migratoria}.
The animal's metathoracic ganglion and nerve were exposed. Spikes
were recorded intracellularly from the axons of auditory receptors
located in the tympanal nerve, see Rokem et al. (2006) for details.
The auditory stimulus was played from a loudspeaker located
ipsilateral to the recorded neurons, at 30 cm from the animal. 37
receptor cells were recorded, from 23 animals. Each cell was tested
with two or more stimuli, resulting in 132 data sets in total (one
data set, or {\em session}, corresponds to one cell in one stimulus
condition). The experimental protocol complied with German law
governing animal care.

\begin{figure}[htdf]
\includegraphics{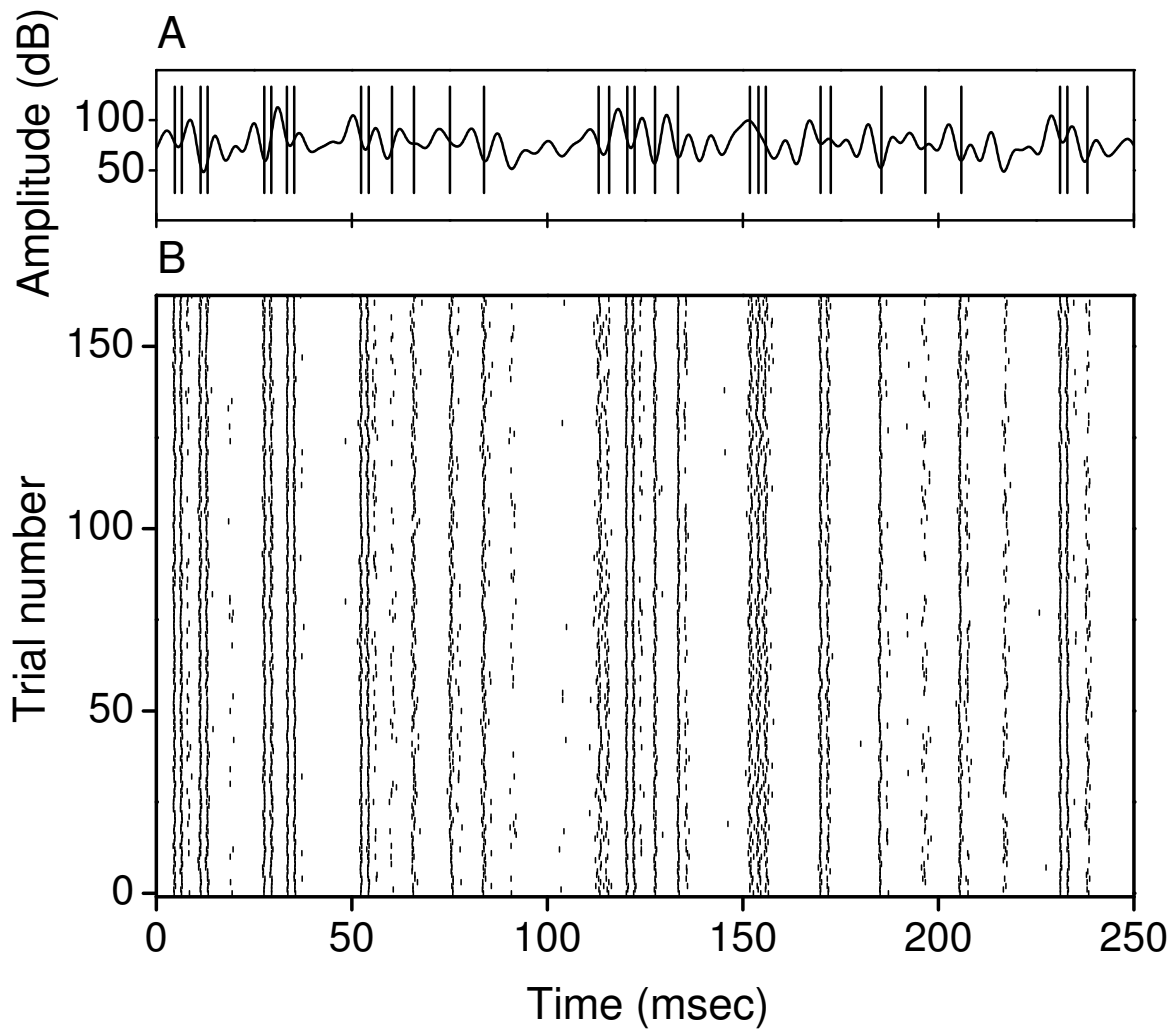}
\caption{\label{f1} Example of an acoustic stimulus and neural
response from a single recording session. {\em A:} Wavy line: random
amplitude modulation (AM signal) of a carrier sine wave. The
standard deviation of the AM signal is 12 dB, its cutoff frequency
is 200 Hz. Vertical lines: elicited spikes. The cell generates
either isolated spikes, or stereotyped patterns consisting of 2-3
spikes separated by a short interval. {\em B:} Raster plot
corresponding to the recording shown in {\em A}, for 165
repetitions. Both the timing of individual spikes and the number of
spikes in each pattern appear as reliable features, fairly well
preserved throughout the different trials.}
\end{figure}

Each experiment began with a measurement of the ``best'' or
``preferred'' sound frequency of the receptor, that is, the
frequency of a sinusoidal acoustic wave for which the threshold of
the cell is lowest. To that end, the animal was exposed to a pure
tone between 3 and 20 kHz. The frequency that induced spiking with
minimal stimulus amplitude was selected as the {\em best frequency}
of the cell, and the minimal intensity inducing spiking constituted
the {\em threshold} $s_{\rm TH}$. The mean threshold across the
population was 58 dB (SD 14 dB). Mimicking behaviorally relevant
stimuli, the sound signals used for further analysis consisted of
amplitude modulated (AM) carrier sine waves whose frequency matched
the cell's best frequency. The AM signal was white up to a certain
cutoff frequency and had a Gaussian amplitude distribution with a
given standard deviation (see Fig.~\ref{f1}, for an example). A
detailed explanation of the stimulus construction may be found in
Machens et al. (2001). Increasing the standard deviation results in
more pronounced variations of the amplitude modulations. By varying
the cutoff frequency, instead, the temporal scale of the stimulus
excursions is altered, with higher cutoff frequencies corresponding
to more rapid amplitude deflections.


Different receptors vary in their cellular properties, resulting in
different response characteristics. To identify the effect of the
stimulus on the response (in spite of the cell-to-cell variability)
each cell was presented with two stimuli. One stimulus was the same
for all cells: a Gaussian amplitude distribution with 6 dB standard
deviation and 200 Hz cutoff frequency. The other signal could be one
of 6 different stimulation protocols. In four of them, the standard
deviation of the amplitude modulation was fixed at 6 dB, and the
cutoff frequency was either 25, 100, 400 or 800 Hz. In the other two
protocols, the cutoff frequency was fixed at 200 Hz, whereas the
standard deviation was set to either 3 or 12 dB.

Given that the mean firing rate has a strong effect on the
transmitted information (Borst and Haag, 2001), the mean stimulus
was adjusted to obtain an average firing rate of about 100 Hz. The
resulting firing rates had a mean of 113 Hz (SD = 16 Hz), and they
did not show any significant variation in the different stimulus
conditions, as assessed by a one-way ANOVA ($p = 0.58$). In
addition, given that information measures require stationary
recordings, we only kept those sessions where the trial-to-trial SD
of the firing rate was lower than 35 Hz (the population average of
this SD is 6 Hz). There were 86 out of 132 data sets that fulfilled
these two conditions.

Once the carrier frequency and mean stimulus amplitudes were
determined, $N$ repetitions of each stimulus were presented, with
$N$ ranging between 98 and 503 (average 172), depending on how long
the recording could be sustained. Each stimulus lasted for one
second, though in all results presented here, the first 200
milliseconds of each trial were discarded, to avoid the initial
transient response, where fast adaptation processes take place.
Different trials were separated by pauses of 700 msec to prevent
slow adaptation effects (Benda and Herz, 2003).

\subsection{Burst identification}
\label{identification}

Neural responses were preprocessed to decide which cells had a
natural tendency to generate bursts, and in these cases, to identify
the bursts. With such a procedure, all spikes should either be
classified as isolated spikes (a 1-spike burst), or be grouped into
bursts of 2 or more discharges (an $n$-spikes burst). We therefore
searched for a reliable criterion to establish a limit value of the
inter-spike interval (ISI) separating pairs of consecutive spikes,
such that all those pairs whose intervals lie below the limit be
considered as part of the same burst, and all those that fall above
the limit be classified as belonging to different bursts. Previous
approaches (see, for example, Kepecs and Lisman, 2003; Metzner et
al., 1998; Oswald et al., 2007; Reich et al., 2000; Reinagel et al.,
1999) have determined the value of the limiting ISI from the shape
of the ISI distribution. In this work, we have taken an alternative
approach, based on the shape of the correlation function.

If a cell shows a tendency to generate bursts, not all intervals
between pairs of spikes are equally probable. We evaluated the
correlation function (also called {\em autocorrelation}) of each
cell discretizing the time axis in $N_{\rm b}$ bins, each of
duration $\delta t = 0.1$ msec. The spike train $\rho(t)$ is
represented as a binary string such that, for any given $t$,
$\rho(t)$ is either equal to $1/\delta t$ or to zero, depending on
whether or not a spike is fired inside $[t, t + \delta t]$. The
post-stimulus-time histogram $r_s(t) = \langle \rho(t) \rangle$ is
the trial average of $\rho(t)$. The mean firing rate $\bar{r}_s =
\sum_t r_s(t) / N_{\rm b}$ is defined as the temporal average of
$r_s(t)$. The correlation function of the spike train is
\begin{equation}
C_s(\tau) = \overline{ \left[\rho(t) - \bar{r}_s\right] \left[
\rho(t + \tau) - \bar{r}_s \right]}, \label{e1}
\end{equation}
where the horizontal bar represents both trial average and temporal
averages over $t$. A large, positive value of $C_s(\tau)$ indicates
that there is a high probability of finding two spikes separated by
a time lag $\tau$, irrespective of whether there are other spikes in
between or not. If $C_s$ is near zero, this probability is roughly
the one to be expected from the mean firing rate of the cell. If
$C_s(\tau)$ is large and negative, the probability that two spikes
be separated by an interval $\tau$ is low.

\begin{figure}[htdf]
\includegraphics[scale = 0.9]{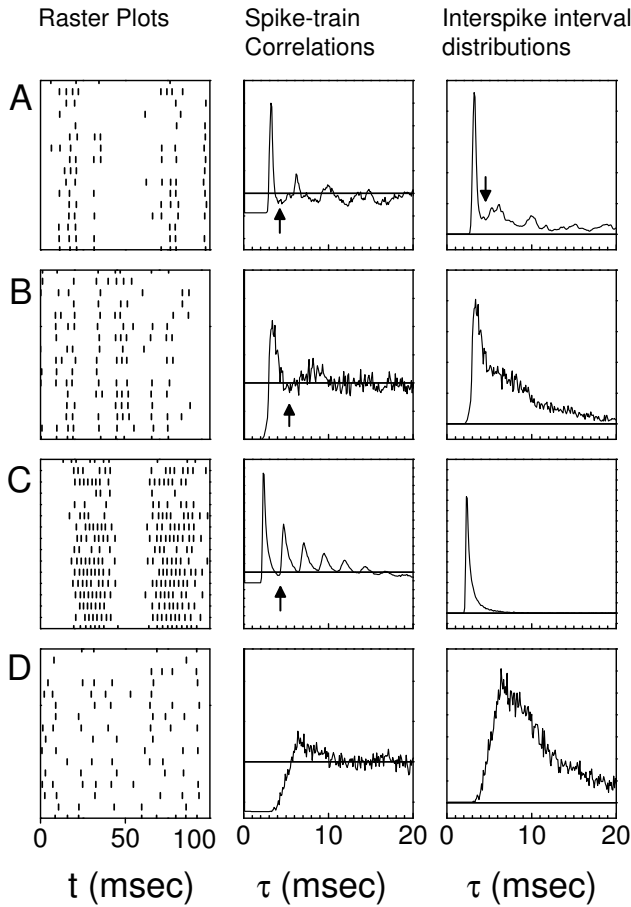}
\caption{\label{f0} Examples of neural responses (left), and the
corresponding spike train correlation functions (middle) and ISI
distributions (right). The four rows of panels depict different
cells. In the middle and right panels, the horizontal line
represents the zero level of the respective quantity. The arrows
indicate the limiting ISI defining burst generation. The upper
three cells ({\em A, B, C}) show a tendency to fire action
potentials separated by a fairly constant ISI, as seen from the
raster plots. The correlation functions allow a clear estimation
of the limiting ISI needed to define bursts, even in cases where
this is not possible using ISI distributions ({\em B} and {\em
C}). The last cell ({\em D}) lacks well defined time scales for
intra-burst and inter-bursts ISIs.}
\end{figure}

Figure~\ref{f0} shows typical responses from four cells. The left
column depicts the response to 15 identical stimulus presentations
to each cell. The correlation functions $C_s(\tau)$ are presented in
the middle column, and for comparison, the ISI distributions
corresponding to the same spike trains are given in the right
column. In cell {\em A}, both the correlation function and the ISI
distribution exhibit a prominent peak. This peak constitutes a clear
signature of the tendency of the cell to fire action potentials
about every 3 msec, as can be seen in the raster plot. The width of
this peak can be easily estimated from either the correlation
function or the ISI distribution, since in both cases the peak is
limited on its right-hand side by a minimum whose location can be
clearly identified (marked by the arrow). In such cases, the
limiting value of the ISI defining burst firing may be set as that
ISI where the minimum is located. However, there are more
complicated cases, too. The following examples ({\em B} and {\em C})
depict two cells that also tend to burst, as shown by the raster
plots. In {\em B}, there are frequent doublets or triplets of
spikes, whereas in {\em C}, each burst typically contains between 6
and 10 spikes. The width of the first peak of the correlation
function can be determined quite easily. However, the temporal span
of the corresponding peak in the ISI distribution is much more
difficult to determine, since the right tail of the peak decreases
essentially monotonically. Moreover, the ISI distribution of cell
{\em C} completely misses the structure of peaks in the
corresponding correlation function.


ISI distributions reflect only the interval between two consecutive
spikes, whereas correlation functions include intervals between any
two spikes. Hence, ISI distributions often show an almost
exponential decay, that conceals some of the structure exhibited by
the correlation functions. For this reason, we shall base our choice
of the limiting ISI defining bursts on the behavior of the
correlation function, and not on the ISI distribution. We have
verified that the two methods give different results only when
applied to cells that have a tendency to generate long bursts
(including more than 5 spikes). In these cases, if our method is
applied to the ISI distributions, it fails to detect the minimum ISI
separating inter-bursts and intra-bursts intervals. The correlation
function, instead, shows a clear multi-peak structure. The example
cell {\em D} is once again simple. It has no tendency to generate
bursts, and consequently, both the correlation function and the ISI
distribution reveal rather broad, unspecific structures.

We stipulated that a cell be classified as {\em bursting} if its
correlation function contained a first peak that was limited on
the right side by a minimum that could be considered significantly
different from the maximum. Below, an {\it ad-hoc} method to
determine the separability of the maximum is provided. In
addition, the maximum was required to lie below $\tau =$ 5 msec,
and the minimum to the right of the maximum should be located
below 1.25 times the inverse cutoff frequency of the AM signal.
These criteria reject fluctuations in the correlation function
arising from limited sampling, as could be any of the many small
troughs observed in Fig.~\ref{f0}{\em B}, and avoid a
misclassification where two consecutive spikes are generated by
two consecutive fluctuations in the stimulus.

To assess whether the correlation function contained a separable
first peak (in the above sense), an {\it ad-hoc} statistical
analysis was performed. To that end, the expected error of the
correlation function was estimated, for all times $\tau$. Notice
that $C_s(\tau)$ can be interpreted as an average (see
Eq.~(\ref{e1})). The error bar $\Delta$ of an average estimated
from $N$ samples reads $\Delta = \sigma / \sqrt{N}$, where
$\sigma$ is the standard deviation of the data to be averaged
(Barlow, 1999). The population mean of the temporal average of
this estimated error was 3.4\% (SD 1.5\%) of the total span of
$C_s(\tau)$ (that is, the difference between the maximum and the
minimum). Two values of $C_s(\tau)$ and $C_s(\tau')$ were
classified as significantly different if they differed in more
than the sum of their estimated error bars. This is an ad-hoc
procedure, since it is based on the assumption that the estimation
errors of $C_{\rm s}(\tau)$ are independent for different times
$\tau$, which may not be the case. However, we have checked that
in all cases, the limiting ISI identified with our method could be
easily detected visually.

Not all cells, and not all stimuli, gave rise to correlation
functions that contained a separable first peak (for example,
Fig.~\ref{f0}{\em D} shows a non-bursting cell). Whenever the peak
could be separated, the {\em domain} of the peak was defined as the
interval between zero and the position of the first minimum after
the peak. In the remaining cases, the domain of the peak was defined
as zero. All spikes in a neural response were assigned to sequences
containing 1, 2, or more action potentials, hereafter called {\em
bursts of intra-burst spike count $n$} or, more compactly, $n$-{\em
bursts}. An $n$-burst was defined as the set of consecutive spikes
whose ISIs fell within the domain of the first peak of the
correlation function. In those sessions where this peak was not
separable, all spikes were classified as 1-bursts, or, as we shall
also call them, as {\em isolated spikes}.

The present method of identifying bursts differs from other criteria
employed previously (Gour\'evitsch and Eggermont, 2007; Kepecs and
Lisman, 2003; Metzner et al., 1998; Oswald et al., 2007; Reich et
al., 2000) in two aspects. First, we use ad-hoc statistical
techniques to prevent small fluctuations, caused by limited
sampling, from hampering burst identification. Second, our approach
is based on the correlation function, and not the ISI distribution.
Both quantities are closely related under various conditions. In
fact, for stationary renewal processes, the correlation function can
be derived through convolution from the ISI distribution (Perkel et
al., 1967). A clear minimum of the correlation function can
therefore be expected if the standard deviation of the ISI
distribution is sufficiently smaller than the mean ISI. On the other
hand, it is more convenient to identify bursting neurons by
analyzing their correlation function. If the minimum in the
correlation function is significant, its location provides the value
of the limiting ISI that is needed to segment a given spike train
into sequences of bursts.

\subsection{Model neurons}

To assess whether complex neural dynamics are needed to
obtain burst-like responses to time-dependent stimuli, we modeled
the firing probability density $r_s(t)$ of a measured cell as a
simple, threshold-linear function of the stimulus, with
added refractoriness, namely
\begin{equation}
r_s(t) = \left\{\int_0^T h(\tau) \left[s^*(t - \tau) -
s^*_0\right] \ d \tau \right\} \ \Theta(t - t_{\rm last} - t_{\rm
ref}), \label{modelo}
\end{equation}
where $s^*(t)$ is defined as
\[
s^*(t) = \left\{
\begin{array}{lll}
s(t) & {\rm if} & s(t) \geq s_{\rm TH} \\ s_{\rm TH} & {\rm if} &
s(t) < s_{\rm TH},
\end{array}
\right.
\]
$s(t)$ is the AM signal extending throughout the interval $[0, T]$,
$s^*_0 = \int_0^T s^*(t) dt / T$ is the temporal mean value of
$s^*(t)$, $h(\tau)$ stands for the filter of the cell, $t_{\rm
last}$ is the time at which the previous spike was fired, $t_{\rm
ref}$ is the refractory period, $s_{\rm TH}$ is the threshold of the
cell, and $\Theta$ is Heaviside step function ($\Theta(t)$ = 0, if
$t < 0$, and $\Theta(t) = 1$, if $t \ge 0$). Note that the stimulus
is thresholded before it is filtered. Gollisch and Herz (2005)
disclosed the detailed processes involved in sound transduction.
They showed that the input current entering the auditory receptor
after acoustic stimulation is a non-linear (quadratic) function of
the sound intensity. Thus, low stimulus amplitudes are ineffective
in generating ionic currents, whereas large intensities have an
amplified effect. In Eq.~(\ref{modelo}), for simplicity, we have
assumed that the non-linearity involved in sound transduction is a
thresholding operation, representing ionic channels that only open
when the AM signal surpasses a certain characteristic value that we
can actually measure. This model, although simplified, correctly
reproduces the threshold-linear dependence of firing frequency vs.
stimulus amplitude that we have observed experimentally for the
stimulus intensities in this study. In Eq.~\ref{modelo}, the current
is further filtered to represent the capacitive properties of the
cell membrane (Gollisch and Herz, 2005). For each modeled cell, the
linear filter $h(\tau)$ was obtained from a cross-correlation
analysis of the spike train and $s^*(t)$ (Koch and Sergev, 1998),
whereas the refractory period $t_{\rm ref}$ was defined as the
minimal ISI of the cell, and $s_{\rm TH}$ was measured
experimentally (see Sect.~\ref{s2p1}). Finally, spike generation was
modeled as a Poisson process with time-dependent rate $r_s(t)$. Note
that the model contains no free fit parameters.

\subsection{Information theoretical analysis}

Brenner et al. (2000) have calculated the mean amount of information
$I_E^{(1)}$ transmitted by an {\em event} $E$, where $E$ is a
pre-defined combination of spikes and silent intervals. Such an
event is either present or absent, in one given trial, at one
particular time. When the event $E$ is a single spike
\begin{equation}
I_E^{(1)} = \int_0^T \frac{r_s(t)}{\bar{r}_s} \ \log_2 \left[
\frac{r_s(t)} {\bar{r}_s} \right] \ {\rm d}t, \label{i1}
\end{equation}
where the event rate $r_s(t)$ is the probability density of a spike
at time $t$ (Brenner et al., 2000; Rieke et al., 1997), and
$\bar{r}_s$ is the temporal average of $r_s(t)$. In Eq.~(\ref{i1}),
the upper index $(1)$ denotes the mean information transmitted by
{\em each} event. Notice that $I_E^{(1)}$ is proportional to the
dissimilarity between the spiking probability density $r_s(t)$ and a
uniform density $\bar{r}_s$, as measured by the Kullback-Leibler
divergence (Cover and Thomas, 1991).

We now extend this analysis to encompass events that are not just
binary (present or absent), but appear in one of several possible
alternatives. In our case, a burst may contain 0, 1, .. or $n$
spikes. For each stimulus stretch $s$ extending during the time
interval $[t - t_0, t]$, the cell generates a response in the time
bin $[t, t + \delta t]$ that may either be ``no spike'' ($n$ = 0),
or the initiation of an $n$-burst $(n > 0)$. The length of the
interval $t_0$ is assumed to be sufficiently large as to contain all
structures in the stimulus that are causally related to the response
of the neuron at time $t$. The mutual information $I^{\delta t}$
between stimuli and $n$-bursts within $[t, t + \delta t]$ is (Cover
and Thomas, 1991)
\begin{equation}
I^{\delta t} = \sum_{s} P(s)\sum_{n = 0}^{+\infty} P(n|s) \log_2
\left [ \frac{P(n|s)}{P(n)} \right] , \label{einfo}
\end{equation}
where $P(s)$ is the prior probability of the stimulus segment $s$,
$P(n|s)$ is the probability of response $n$ whose first spike falls
in the interval $[t, t + \delta t]$ conditional to the stimulus $s$,
and
\begin{equation}
P(n) = \sum_{s} P(n|s) P(s) \label{prior}
\end{equation}
is the prior probability of response $n$. In Eqs.~(\ref{einfo}) and
(\ref{prior}) the sums in $s$ include all possible stimulus
stretches spanning the interval $[t - t_0, t]$, each one of them
with its probability $P(s)$.

If $\delta t$ is sufficiently small, then for all $n > 0$ the
probability $P(n|s)$ may be approximated by $r_n(s) \delta t$, where
$r_n(s)$ is the $n$-burst rate conditional on the stimulus $s$, and
is proportional to the fraction of trials where an $n$-burst was
initiated in $[t, t + \delta t]$, in response to stimulus $s$.
Similarly, $P(0|s) \approx 1 - \delta t \sum_{n = 1}^{+ \infty}
r_n(s)$. Replacing these expressions in Eq.~(\ref{einfo}) results in
\[ I^{\delta t}
\approx \delta t \sum_{s} P(s) \sum_{n = 0}^{+\infty} r_n(s)
\log\left[\frac{r_n(s)}{\bar{r}_n}\right], \] where
\begin{equation}
\bar{r}_n = \sum_s P(s) r_n(s).
\end{equation}
If the stimulus is stationary, all possible stimulus stretches $s$
will eventually be found as time goes by, each one of them with a
frequency that is proportional to $P(s)$. Therefore, for long enough
stimuli, averaging over $s$ with the probability distribution $P(s)$
may be replaced by time averaging. That is,
\begin{equation}
I^{\delta t} \approx \frac{\delta t}{T} \ \sum_{n = 0}^{+\infty} \
\int_0^T  r_n(t) \log \left[ \frac{r_n(t)}{\bar{r}_n} \right] {\rm
d}t , \label{einfo3}
\end{equation}
where now the $n$-burst rate $r_n(t)$ is expressed as a function of
time, and
\[
\bar{r}_n = \frac{1}{T} \int_0^T r_n(t) \ {\rm d}t.
\]

Equation (\ref{einfo3}) provides a first estimate of the mutual
information between stimuli and responses in a short interval $[t,
t + \delta t]$. The aim is now to extend this result to the whole
response interval [0, $T$], which can be thought of a
concatenation of small intervals $[0, \delta t], [\delta t, 2
\delta t],$ ... $[(k - 1) \delta t, k \delta t]$, where $k = T /
\delta t$. This extension, however, can only be done if the
response in one time interval does not depend on the response in
another time interval. Consider the response vector $\vec{n}(t) =
(n(t), n(t + \delta t), n(t + 2 \delta t), ..., n[t + (k - 1)
\delta t])$, where $n(\tau)$ represents the number of spikes
contained in the burst whose first spike fell in $[\tau, \tau +
\delta t]$ ($n = 0$ means that the cell remained silent). If
different time bins are independent, then
\begin{equation}
P[\vec{n}(t)] = \Pi_i P[n(t + i\delta t)]. \label{indep1}
\end{equation}
This means that that responses in different time bins are
independent from one another, given a fixed stimulus history. Full
independence of time bins, however, implies that the factorization
of Eq.~(\ref{indep1}) should not only hold for each stimulus
history, but also for the marginal probabilities
\[
P(\vec{n}) = \frac{1}{T} \ \int_0^T P[\vec{n}(t)] \ {\rm d}t, \ \
\ \ \ \ \ {\rm and} \ \ \ \ \ \ \ P[n(i\delta t)] = \frac{1}{T} \
\int_0^T P[n(t + i \delta t)] \ {\rm d}t.
\]
These quantities represent the probability of the word $\vec{n}$ and
the $i$-th bit $n$ inside the word at {\em any} temporal location
within the spike train. Then, if different time bins are
independent, in addition to Eq.~(\ref{indep1}), we must also have
\begin{equation} P(\vec{n}) = \Pi_i P[n(i\delta t)],
\label{indep2}
\end{equation}
implying that independence also holds for arbitrary stimulus
histories. When these two conditions are fulfilled, and given the
additive properties of information (Cover and Thomas, 1991), the
mutual information $I$ between stimuli and responses in $[0, T]$ is
the sum of the mutual information between stimuli and responses in
each sub-interval $[(j - 1) \delta t, j \delta t]$. Hence,
\begin{equation}
I = k \times \ I^{\delta t} = \sum_{n = 0}^{+ \infty} \int_0^T
r_n(t) \log_2 \left[\frac{r_n(t)}{\bar{r}_n} \right] {\rm d} t
\equiv \sum_{n = 0}^{+ \infty} \bar{r}_n I^{(1)}_n \equiv \sum_{n =
0}^{+\infty} I_n,\label{efinal}
\end{equation}
where the last two equivalences serve as definitions of the average
information $I_n^{(1)}$ transmitted by each single $n$-burst, and
the information $I_n$ transmitted by all the bursts of a given $n$,
respectively. Finally, the information per unit time $I'$ (also
called {\em information rate}), and the rates $I'_n$ are obtained by
dividing the corresponding expressions in Eq.~(\ref{efinal}) by the
total time interval $T$.

We emphasize that Eq.~(\ref{efinal}) is only valid under the
independence assumption, that is, if Eqs. (\ref{indep1}) and
(\ref{indep2}) hold. In this work, we assume that all correlations
in the spike train of third or higher order can be neglected. Under
this approximation, different time bins are independent, if they are
uncorrelated. This means that the probability distribution of a
binary string $\vec{n} = (n_1, ..., n_k)^T$ is well approximated by
a Gaussian function $P(\vec{n}) = \exp\left[-(\vec{n} - \langle
\vec{n} \rangle)^T \Sigma^{-1} (\vec{n} - \langle \vec{n} \rangle)/2
\right] / \sqrt{(2 \pi)^k \det \Sigma}$, where $\Sigma_{ij} =
\langle (n_i - \langle n_i \rangle)(n_j - \langle n_j \rangle)
\rangle$. This approximation should hold both for strings $\vec{n}$
starting at a fixed time $t$, and also for any time. The Pearson
correlation coefficient between time bins
\begin{equation}
c_{b}(t, \tau) = \frac{ \Big \langle \ n^*(t) \ \  n^*(t + \tau)\
\Big \rangle } {\left \langle \left[\ n^*(t)\ \right]^2 \right
\rangle^{1/2} \ \left \langle \left[\ n^*(t + \tau)\ \right]^2
\right \rangle^{1/2} } \label{cb}
\end{equation}
quantifies the correlations between $n(t)$ and $n(t + \tau)$ for a
fixed stimulus history, and hence may be used to test whether
Eq.~(\ref{indep1}) is valid. In Eq.~(\ref{cb}), $n^*(t)=n(t) -
\langle n(t) \rangle$, and the angular brackets represent trial
averages. In order to make Eq.~(\ref{cb}) well defined even at times
when the response of the neuron has no variability (that is, $\left
\langle \left [ n^*\left(t\right)\right]^2\right\rangle =0$ or
$\left \langle \left [ n^*\left(t+\tau\right)\right]^2\right\rangle
=0$), we set $c_b(t, \tau) \equiv 0$ if both the numerator and the
denominator vanish.

In the absence of higher-order correlations, whenever $c_b(t, \tau)
\approx 0$ for all $t$ and $\tau$, one can assert that
Eq.~(\ref{indep1}) holds. To assess whether burst identification
succeeded in decreasing the correlations in the spike train, $c_b(t,
\tau)$ should be compared with a similar correlation coefficient
$c_s(t, \tau)$ calculated from a binary representation of the spike
train including the whole collection of spikes. $c_s(t, \tau)$ is
defined by a formula analogous to Eq.~(\ref{cb}), but with the
integer variable $n$ replaced by a binary variable indicating the
presence or absence of a spike in each time bin. To quantify the
total amount of correlations in a given domain $t \in [t_1, t_2]$
and $\tau \in [\tau, \tau']$, we use the mean square value of the
Pearson correlation coefficient ($c_b(t, \tau)$ or $c_s(t, \tau)$)
in the selected domain.

The Pearson correlation coefficient between $n(t)$ and $n(t + \tau)$
for any stimulus history is
\begin{equation}
c_b(\tau) = \frac{\overline{^{\phantom{2}}\left[\ n^*(t)\ \right]
\left[\  n^*(t + \tau) \ \right]^{\phantom{2}}}}{\Bigg\{
\overline{\left[\ n^*(t)\ \right]^2} \ \ \overline{\left[\ n^*(t +
\tau)\ \right]^2} \Bigg\}^{1/2}}, \label{corr}
\end{equation}
where the bar represents both a trial and a temporal ($t$) average.
In the absence of higher-order correlations, whenever $c_b(\tau)
\approx 0$ for all $\tau$, one can assert that Eq.~(\ref{indep2})
holds. To compare the correlations between bursts with the
correlations between spikes, Eq.~(\ref{corr}) should be compared
with $c_s(\tau)$, defined by a formula analogous to
Eq.~(\ref{corr}), but with the integer variable $n$ replaced by a
binary variable representing individual spikes.

\subsection{Estimation of burst-triggered averages}

The spike-triggered average (STA) was calculated as the mean
stimulus preceding a spike, namely,
\[
{\rm STA}(\tau) = \frac{1}{N_0} \sum_{t_0} s(t_0 + \tau),
\]
where $s(t)$ is the time-dependent stimulus, $N_0$ is the total
number of spikes, and the sum ranges over all spike times $t_0$. In
every investigated cell, STA$(\tau)$ showed a pronounced peak. The
time between the maximum of the peak and $\tau = 0$ (spike
generation) is the average latency between upward stimulus
deflections and spike occurrences. As an extension, the $n$-burst
triggered averages ($n$BTAs) were introduced to represent the mean
stimulus preceding an $n$-burst (Kepecs and Lisman, 2003; Lesica et
al., 2006; Oswald et al., 2007), that is,
\begin{equation}
n{\rm BTA}(\tau) = \frac{1}{N_n} \sum_{t_n} s(t_n + \tau),
\label{nbta}
\end{equation}
where now, the sum ranges over all times $t_n$ at which an $n$-burst
begins (that is, the time of the first spike), and $N_n$ is the
total number of $n$-bursts. The time $\tau_n$ between the maximum of
$n{\rm BTA}$ and $\tau = 0$ (burst generation) is the average
latency of the $n$-burst.

The $n$BTA at a particular $\tau$ is the arithmetical average of a
collection of values, whose standard deviation reads
\begin{equation}
\sigma_{n}(\tau)  = \sqrt{\frac{1}{N_n - 1} \sum_{t_n} \left[s(t_n +
\tau) - n{\rm BTA}(\tau)\right]^2}. \label{snbta}
\end{equation}
To determine whether the $n$BTAs corresponding to different $n$
values differed significantly, an ANOVA was conducted. The test was
performed in the frequency domain, to avoid temporal correlations.
The $n$BTA in the time interval ranging from -25 to +15 msec from
burst generation was Fourier transformed and a two-way ANOVA was
separately conducted on the real and imaginary parts of the
frequency representation of the signal (since these constitute two
comparisons, Bonferroni's correction for multiple hypothesis testing
was incorporated), with frequency band and the order of the burst as
factors in the analysis. The null hypothesis was 1BTA = 2BTA = 3BTA
= 4BTA. The corrected significance level was set at 0.01. Cells
showing a significant difference (either as a main effect, or an
interaction) were further tested in the time domain, to determine
the intervals where the difference was observed. This was done using
independent t-tests, for each point in time. In this case, the null
hypothesis was that at time $t$, $n$BTA$(t)$ differed from at least
one of the other $n'$BTA$(t)$, for any $n'\ne n$. In this analysis,
$n$ and $n'$ ranged between 1 and 4. Hence, to reject the null
hypothesis for a given $n$ and $t$, 3 comparisons with different
$n'$ values are needed.

For $n \ge 2$, we also compared the $n$BTAs with a combination of
$n$ 1BTAs interleaved with the same ISIs found in the real data. For
every $n$-burst in the experimental data, we calculated the function
\begin{equation}
f_n(t) = \sum_{i = 1}^n 1{\rm BTA}(t - t_i), \label{convolved}
\end{equation}
where the times $t_i$ indicate the location of each spike within the
burst. Each $n$-burst, hence, produces a function $f_n(t)$. By
averaging the $f_n(t)$ obtained for all bursts with the same spike
count $n$, we calculated the averaged convolved 1BTA. We estimated
the variability of the convolved 1BTA as the standard deviation of
the averaged data. To test whether the real $n$BTA was significantly
different from the reconstructed $f_n$, we first carried out a
two-way ANOVA. The null hypothesis was $n$BTA = $f_n$ in a time
interval extending between the two minima at each side of the
central maximum of the $n$BTA. To avoid temporal correlations, the
comparisons were performed in Fourier space, testing real and
imaginary parts separately. A Bonferroni correction for multiple
comparisons was incorporated. The corrected significance level was
set at 0.01. Cells showing a significant difference (either as a
main effect or an interaction) where further tested in the time
domain, to determine whether the difference was observed in an
extended fraction of the time interval. This was done with an
independent t-test, for each point in time. In this case, the null
hypothesis was that at time $t$, $n$BTA$(t) = f_n(t)$. We reported
the number of cells for which the null hypothesis was rejected in
70\% of the times $t$ within an interval extending between the two
minima at each side of the central maximum of the $n$BTA. As a
check, the whole procedure was also carried out replacing the
1BTA$(t)$ in Eq.~(\ref{convolved}) with STA$(t)$. Recall that the
1BTA is the average stimulus preceding 1-bursts, or isolated spikes.
The STA, in turn, is the average stimulus preceding all action
potentials in the spike train.

For completeness, we mention that the amount of jitter (Rokem et
al., 2006) is defined as the trial-to-trial standard deviation of
the time of the first spike in a burst, and the average estimated
error bar in jitter estimation is 0.2 msec.

\subsection{Relating burst probabilities to the height of stimulus
excursions}

To calculate the probability $P(n | h)$ of obtaining a burst with
$n$ spikes after a stimulus deflection of maximal height $h$, we
went through all local maxima of the stimulus, one at a time, and
for each one we searched whether there was a burst in the response
that could be associated with the maximum. This was done in the
following way. Each $n$-burst in the response was first shifted
backwards $\tau_n$ milliseconds. Next, for a given stimulus maximum
located at time $t_0$, we searched for (shifted) $n$-bursts inside a
window $[t_0 - T, t_0 + T]$, where $T$ was the width of the most
prominent peak of the STA of the whole collection of spikes (prior
to burst identification). In other words, $T$ was the interval where
a given response can be expected to be correlated with a maximum in
the stimulus. If within that interval no bursts were found, then the
maximum located at $t_0$ was said not to be associated with any
response. If the first spike of an $n$-burst fell within the window,
then the maximum in the stimulus was associated with that $n$-burst.
If there was more than one burst inside the window, then a single
burst was selected, by choosing that one whose first spike lay
closest to $t_0$. Next, if a given burst was associated to more than
a single maximum, the closest maximum was assigned to the burst (and
not the others).

This algorithm allows one to associate each maximum in the stimulus
with either no response, or with an $n$-burst. Note, however,
that so far we have no reason to claim that there is a
causal connection between the maximum and the associated burst. In
principle, given that we do not actually know what feature in the
stimulus induces burst generation (it could be the height of the
stimulus amplitude, the size of its derivative, the width of an
upward excursion, and so forth) this association between stimuli and
responses could represent no more than a completely arbitrary
connection. Only if we can show that the association contains
non-trivial features that would be unlikely between randomly
connected events can we suspect that it could indeed contain some
predictive value.

To reveal those features, we estimated
$P(n|h \in [h_0 - \Delta h, h_0 + \Delta h])$, i.e., the
probability of obtaining a burst of $n$ spikes, given that the
height of the stimulus maximum $h$ fell in $[h_0 - \Delta h, h_0 +
\Delta h]$. The width $\Delta h$ was chosen as $5\%$ of the span of
values of $h$. $P(n|h)$ is depicted in Fig.~\ref{probability} for
an example cell. The partial segregation between the different
curves shows that the height of the maximum $h$ can tell something
about the stimulus. Even though one still cannot guarantee a causal
relationship between each maximum and its associated $n$-burst, this
result ensures that the intra-burst spike count $n$ provides
information about the height of the stimulus deflection preceding it
- not excluding that it may also provide information about other
stimulus features.


\section{Results}

\subsection{Stimulus characteristics modulate burst probability}

\label{stimulus}

Depending on the characteristics of the ionic channels that
compose the cellular membrane and temporal properties of their
activation and inactivation variables, different neurons respond
to the same stimulus with different firing patterns. In
particular, some neurons have a tendency to alternate between
periods of high-frequency discharges and silent intervals. This is
called burst firing. The mathematics of burst firing has been
studied extensively in the computational neuroscience literature
(see, for example, Izhikevich, 2000; Izhikevich and Hoppensteadt
2004; Wang and Rinzel, 1995). Irrespective of the particular
mechanisms underlying the generation of bursts, here we explore
their role in the transmission of sensory information. To that
end, we quantify the reliability with which bursts correspond to
specific stimulus features.

In principle, the possibility to generate bursts would allow a
neuron to construct a non-trivial temporal code, in which both the
time at which the burst initiates and the number of spikes within a
burst carry specific information. In order to assess whether this is
the case in a classic insect model system (Gollisch and Herz, 2005;
Hill, 1983; Machens et al., 2001, 2005; R\"omer, 1976; Ronacher and
R\"omer, 1985; Sippel and Breckow, 1983; von Helversen and von
Helversen, 1994), the activity of grasshopper auditory receptor
neurons was recorded {\em in vivo} during acoustic stimulation.
Figure~\ref{f1}{\em A} depicts an example stimulus (wavy line),
together with the elicited spikes (vertical lines). This cell
sometimes generates isolated action potentials, whereas at other
times it fires spike doublets or triplets. In this particular
recording, responses typically appear after stimulus upstrokes with
an delay of ~3.4 msec, including both acoustic and axonal time lags.
The data suggest that whereas fairly shallow stimulus excursions are
followed by, at most, a single action potential, deflections that
are more pronounced (either in height or in width) are often
accompanied by short sequences of multiple spikes.
Figure~\ref{f1}{\em B} depicts the response of the same neuron to
165 identical repetitions of the stimulus. Clearly, the bursting
pattern of this cell is highly reproducible across trials.

\begin{figure}[htdf!]
\includegraphics{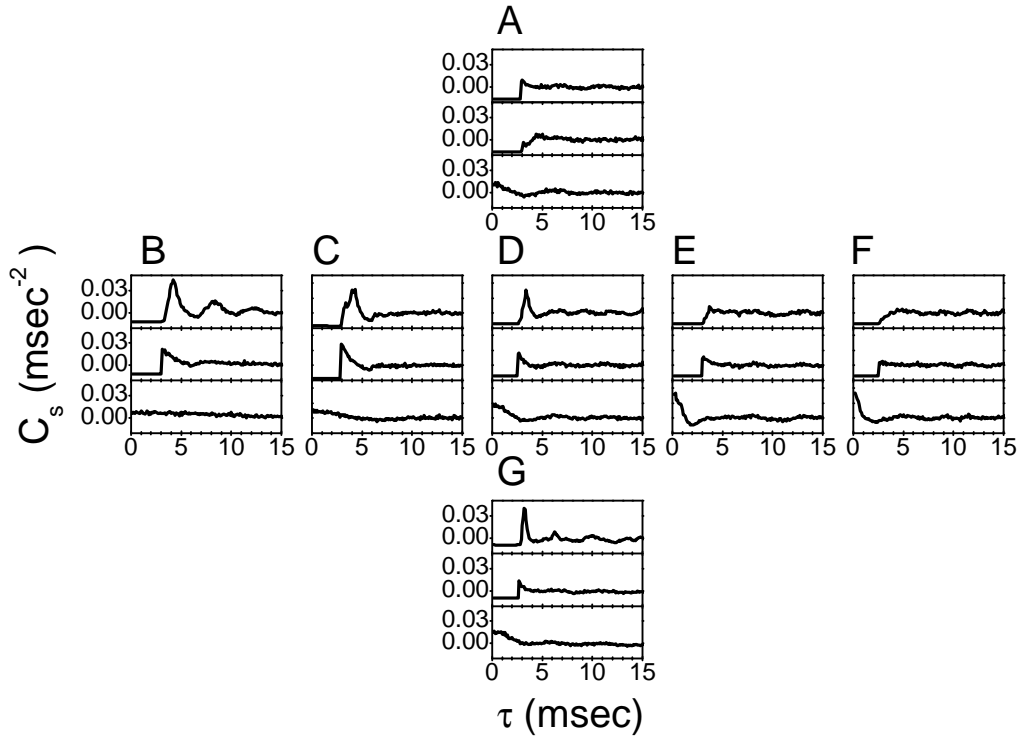}
\caption{\label{f2} Spike-train correlations for a sample cell, and
different stimulus conditions. Each sound stimulus consisted of a
carrier wave with random Gaussian amplitude modulations that had a
specific standard deviation and cutoff frequency. {\em Upper
subpanel}: Experimental data. {\em Middle subpanel:}
Threshold-linear model, with refractory period. {\em Lower
subpanels:} Linear model. Neither model contains free fit
parameters. Comparisons between the experimental data and the two
models demonstrate that the combination of threshold and
refractoriness captures the qualitative shape of the measured
correlation functions. {\em A, D, G:} cutoff frequency = 200 Hz, and
standard deviation 3 dB ({\em A}), 6 dB ({\em D}) and 12 dB ({\em
G}). {\em B} to {\em F:} standard deviation = 6 dB, and cutoff
frequency 25 Hz ({\em B}), 100 Hz ({\em C}), 200 Hz ({\em D}), 400
Hz ({\em E}), and 800 Hz ({\em F}).}
\end{figure}

These observations suggest that short sequences of high-frequency
firing appear with higher probability in response to particular
types of stimulus deflections. This raises the question whether the
probability of generating bursts depends on the statistical
properties of the sound wave. We therefore calculated the
correlation function $C_s(\tau)$ of the neural response (see
Methods). The upper subpanels of Fig.~\ref{f2} show $C_s$ for a
sample cell that was tested with the whole set of stimuli (the
middle and lower subpanels correspond to simulated data discussed
later on). Increasing the standard deviation of the amplitude
distribution (from {\em A} to {\em D} to {\em G}) results in
correlation functions that exhibit progressively sharper peaks. This
is the signature of a high probability of generating sequences of
two or more spikes separated by a fairly constant ISI. Moreover, a
somewhat rippled pattern can be observed in the right tail of the
distribution in {\em G}. Decreasing the typical time scale of the
stimulus fluctuations (going right from {\em B} to {\em F}) leads
from multi-modal ({\em B}) to single-peaked ({\em C}, {\em D}) to
increasingly shallower and broader correlation functions ({\em E},
{\em F}).


Some correlation functions exhibit a pronounced first peak, easily
distinguishable from the rest of the function (as in
Fig.~\ref{f2}{\em B}, {\em C}, {\em D} and {\em G}), and spanning a
finite and fairly clear temporal domain. In these cases, spikes are
either closely packed with ISIs falling in the domain covered by the
first peak, or they are loosely spread apart. The presence of a
minimum between the first peak and the rest of the correlation
function allows one to establish a natural upper limit to the range
of preferred ISIs. Sometimes, this minimum is also present in the
ISI distribution. In these cases, the cell has a tendency to fire
with a typical ``short'' ISI that is clearly separated from other
long ISIs. If the minimum only appears in the correlation function,
but not in the ISI distribution, then the separation between these
two time-scales cannot be achieved directly using the ISI
distribution (see Methods). However, the tendency of the cell to
fire sequences of 3 or more spikes with one typical ISI can still be
clearly revealed by the correlation function. Finally, there are yet
other cases where the correlation function is of an essentially
unimodal nature, exhibiting no more than one broad, unspecific
structure (Fig.~\ref{f2}{\em A}, {\em E} and {\em F}). In these
cases, singling out a range of ISIs as ``typical'' would be
questionable.

We define a {\em burst} as a sequence of spikes whose ISIs fall
within the domain of the first peak of the correlation
function, whenever such peak can be isolated (see Methods, for the
statistical techniques used to assess the separability of this
peak). This sequence of $n$ spikes will be called {\em a burst of
intra-burst spike count} $n$ or, more compactly, an $n$-{\em burst}.
In what follows, the temporal location of a burst is assigned to the
time when its first spike occurs. Cells showing unimodal correlation
functions are classified as non-bursting, and in the analysis below,
all their spikes are considered as 1-bursts.

\begin{figure}[htdf!]
\includegraphics{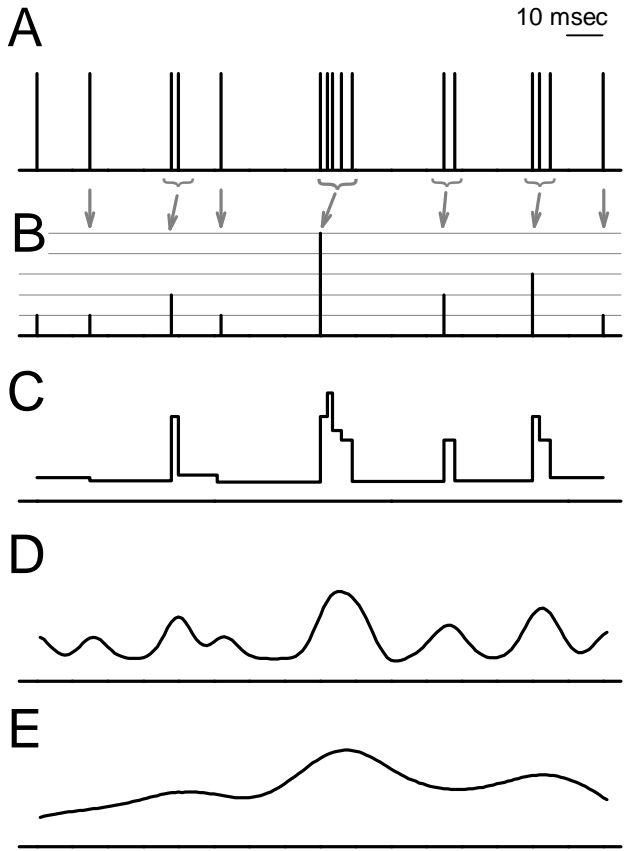}
\caption{\label{fcodes} Graphical representation of different coding
schemes. {\em A}: Sample spike train. For this example, all
consecutive spikes separated by less than 3 msec are considered as
part of the same burst. {\em B}: $n$-burst representation of the
spike train. Each point in time $t$ is associated with an integer
$n$ representing the number of spikes in a burst (if any) initiated
at $t$. The height of the vertical lines represents $n$, and the
arrows indicate the association between each burst in {\em A} and
the corresponding $n$ value in {\em B}. {\em C}: Instantaneous
firing rates, defined as inverse ISIs. {\em D}: Smoothed-firing-rate
representation, defined as the convolution of the spike train with a
Gaussian function of 5 msec SD. {\em E}: same as {\em D}, but using
a Gaussian function of 20 msec SD. Unlike traditional firing-rate
codes ({\em C-E}), the $n$-burst code provides a reduced
representation of the spike train - all ISIs shorter than the ISI
cutoff used for burst definition are treated equally. In addition,
the number of spikes in a burst can be directly read off from the
$n$-burst representation whereas it is not locally available within
firing-rate codes.}
\end{figure}

To underscore the differences between the $n$-burst code
investigated in this study and the more conventional firing-rate
codes, Fig.~\ref{fcodes} illustrates alternative representations of
a sample spike train. Here, {\em rate code} is used whenever the
stimulus is encoded by the firing rate, which is
evaluated either instantaneously (as in {\em C}), or in
extended time windows ({\em D} and {\em E}) . In {\em A}, each
vertical line represents an action potential of a cell that tends to
generate high-frequency bursts with intra-burst ISIs of 2-3 msec.
Panel {\em B} depicts the $n$-burst representation of this spike
train. Here, each time $t$ is associated with an integer $n$ that
denotes the number of spikes contained in the burst starting at time
$t$. The height of the vertical lines in {\em B} represents the
value of $n$, and the grey arrows link each burst in {\em A} with
the corresponding $n$-value in {\em B}. For comparison, three
firing-rate codes are shown in {\em C-E}. Panel {\em C} illustrates
the time-dependent instantaneous firing rate which is obtained from
the sequence of inverse ISIs. Panels {\em D} and {\em E} depict two
alternative smoothed firing-rate representation. In {\em D}, each
spike from {\em A} was convolved with a narrow bell-shaped kernel
(Gaussian, 5 msec SD); in {\em E}, the SD is 20 msec.


For invertible kernels, the firing-rate representations of
Fig.~\ref{fcodes}{\em C-E} contain all information needed to
reconstruct the full spike train in {\em A}. This is clearly {\em
not} the case for the $n$-burst representation in {\em B}. Here,
small variations of the intra-burst ISIs in {\em A} are no longer
present. On the other hand, the number of spikes within a burst
provided by the $n$-burst code is not locally available from the
firing rate-codes in Fig.~\ref{fcodes}{\em C-E}. For these two
reasons, the $n$-burst code is qualitatively different from a
firing-rate code. The reduced information capacity of the $n$-burst
code could severely limit its potential role for neural systems. It
may, however, also provide a highly compact and thus most useful
neural code. The present study aims at elucidating these
alternatives.

Table \ref{t1a} lists all stimulation protocols, together with a
summary of the bursting properties of the investigated cell
population. The fraction of bursting sessions, the percentage of
isolated spikes (1-bursts), and the maximum $n$ value depend
strongly on the standard deviation and cutoff frequency of the
stimulus. Notice, however, that in all cases, isolated spikes are
more frequent than any other burst of $n > 1$.


\begin{table}[htdf!]
\includegraphics{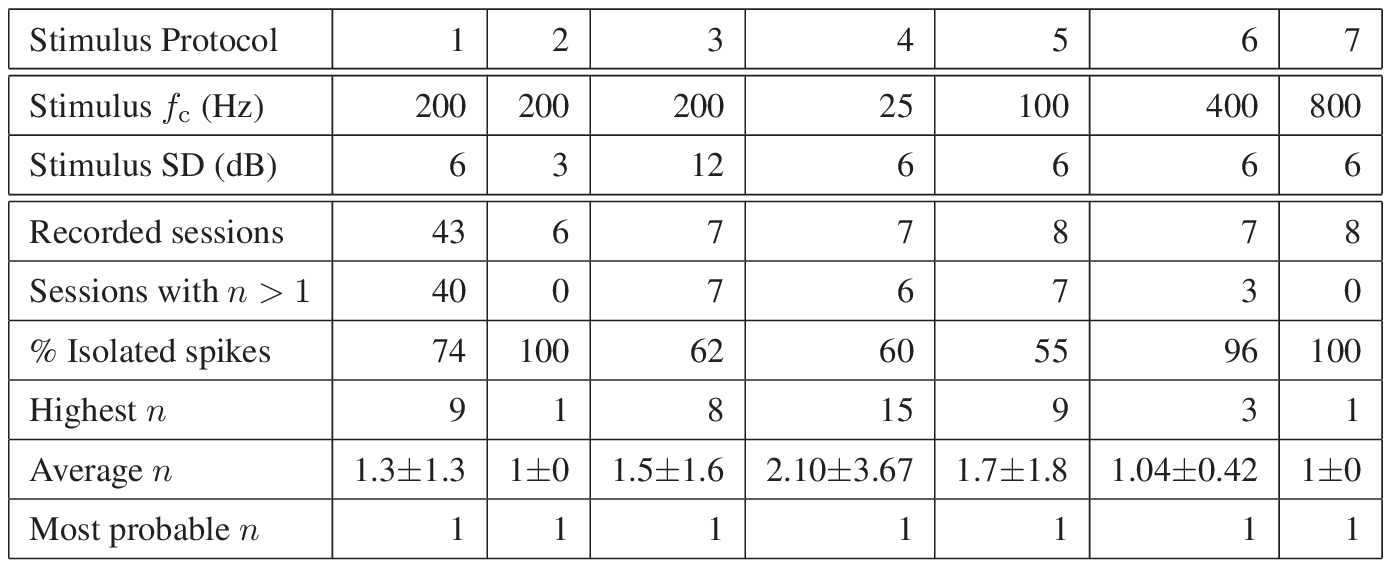}
\caption{Summary of the recorded data. Each column represents a
different stimulation protocol. {\em Stimulus} $f_{\rm c}$: cutoff
frequency of the AM signal. {\em Stimulus SD:} standard deviation of
the AM signal. {\em Recorded sessions:} number of data sets with
that particular protocol. {\em Sessions with $n > 1$}: number of
sessions with bursts with $n > 1$. \% {\em Isolated spikes:} ratio
of the number of 1-bursts to the total number of bursts, in all
bursting sessions. {\em Highest} $n$: highest value of $n$. {\em
Average} $n$: All bursting sessions are pooled together, and for
each $n$, the ratio of the number of $n$-bursts to the total burst
number is calculated. This ratio serves as an estimation of the
probability of finding a given $n$ value. With this probability, the
average $n$-value is estimated, and presented together with its
standard deviation. {\em Most probable} $n$: the $n$ value with
highest probability. } \label{t1a}
\end{table}


Different cells have different firing thresholds, and may therefore
respond to the same stimulus with different mean firing
rates. Both the burst statistics and the transmitted information
depend on the firing rate. In order to be able to compare the
results obtained for different cells, in all experiments reported
here the mean stimulus amplitude was adjusted so as to obtain a mean
firing rate near 100 Hz (see Methods). We also
checked that the firing rate practically has no effect on the value
of the limiting ISI defining bursts. More specifically, a 50 Hz
increase in firing rate shifts the limiting ISI by less than 0.4
msec, which is comparable to its estimated error bar. The
average intra-burst spike count $n$, in turn, shows an increase of
less than 25\%.

\begin{figure}[htdf!]
\includegraphics{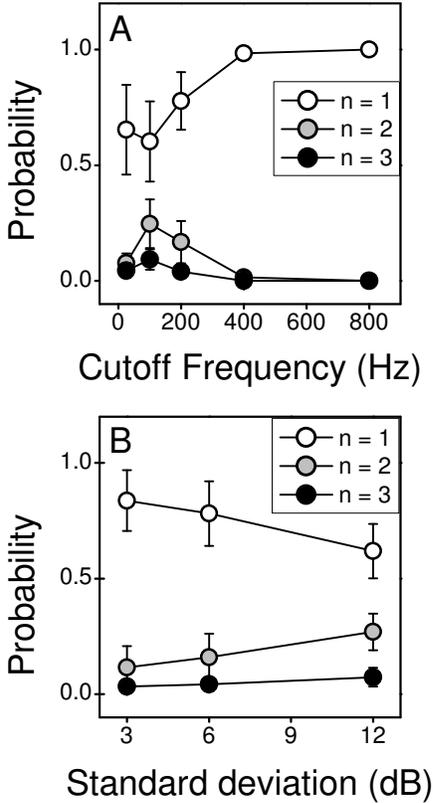}
\caption{\label{f3} Population average of the probability of
generating $n$-bursts, as a function of the stimulus cutoff
frequency, for all SD = 6 dB stimuli ({\em A}) and as a function of
standard deviation, for all stimuli with a cutoff frequency of 200
Hz ({\em B}).  Error bars represent the standard deviation in the
population. High-$n$ bursts appear most frequently for stimuli with
large amplitude modulations and cutoff frequencies around 100 Hz.
Stimulus properties thus have a noticeable influence on the
probability of generating bursts.}
\end{figure}

Stimulus statistics strongly influence the probability of generating
specific bursts, as shown in Fig.~\ref{f3}. Here, the probability of
an $n$-burst is depicted as a function of the cutoff frequency of
the AM signal ({\em A}) and its standard deviation ({\em B}). The
probability of generating isolated spikes is minimal for large
amplitude fluctuations and cutoff frequencies around 100 Hz.  For
the sake of clarity, only data corresponding to $n = 1$, 2 and 3 are
depicted.


In the present approach, a spike sequence is classified as an
$n$-burst by analyzing the statistical properties of the response.
There are no dynamical explanations in terms of specific ionic
currents. Actually, though we lack a detailed characterization of
the ionic currents involved in action potential generation, previous
studies suggest that grasshopper receptors do not burst
intrinsically; cells fire tonically for time-independent stimuli
(Gollisch et al., 2002) and do not show burst activity at the onset
of step-like stimuli (Gollisch and Herz, 2004). In addition,
adaptation effects as well as spike-time variability can be
explained on a quantitative level with models that do not contain
intrinsic burst mechanisms (Benda et al., 2001; Gollisch and Herz,
2004; Schaette et al., 2005). These results underscore that in the
presence of time-dependent stimuli, even cells that do not burst by
themselves may generate responses whose statistical properties are
highly reminiscent of intrinsically bursting cells. Ag\"uera y Arcas
et al. (2003) and Keat et al. (2001) present similar examples in
simulated data. In these cases, burst-like responses arise as a
consequence of the interplay between the dynamical properties of the
neuron and particular temporal structures in the stimulus. To assess
whether even cells with very simple dynamics can exhibit burst
activity when driven by the proper stimulus, we modeled the time
evolution of a threshold-linear Poisson neuron with added
refractoriness (see Methods).  The middle subpanels of Fig.~\ref{f2}
depict the correlation functions for a model cell with the same
filter characteristics, threshold and refractory period as the data
shown in the upper subpanels (see Methods). These correlation
functions exhibit similar qualitative features as those of the real
cell. Recall that the modeled cells contain no free fit parameters.
In both real (upper subpanels) and simulated (middle subpanels)
data, the sessions that are classified as bursting (or non bursting)
coincide. When the analysis is extended to the whole population of
cells, this agreement is observed in 86\% of all sessions. Moreover,
in those sessions where both real and simulated data are classified
as bursting, the limiting ISI calculated with real and simulated
data differ by less than 1 msec in 81\% of the cases. However, the
multiple peaks typically caused by slow stimuli (see, e.g., {\em B})
are only partially reproduced, indicating that the high temporal
precision of subsequent spikes in multiple bursts is not captured by
the simulations. Notice that refractoriness needs to be included in
the model, otherwise the first peak in the correlation function
shifts to $\tau = 0$. Moreover, if the stimulus is not thresholded,
the statistics of the modeled cell differs markedly from the real
one. This is shown in the lower subpanels of Fig.~{\ref{f2}}, where
the correlation function of a purely linear model with the same
filter characteristics as the real cell is depicted. This model
completely fails to capture the basic statistics of the experimental
data, as can be judged from the absence of both the refractory
period and the sharp peak in the correlation function.

\subsection{Quantitative description of the information transmitted by
bursts}

\label{sinfo}

Since the stimulus characteristics have a strong effect on the
probability of burst generation, the number of spikes in a burst may
encode specific stimulus aspects. If this hypothesis is indeed true,
even a reduced burst representation of the spike train should carry
information about the stimulating sound wave. The purpose of the
present section is to translate this general idea into a
quantitative information-theoretical analysis.

\begin{figure}[htdf!]
\includegraphics{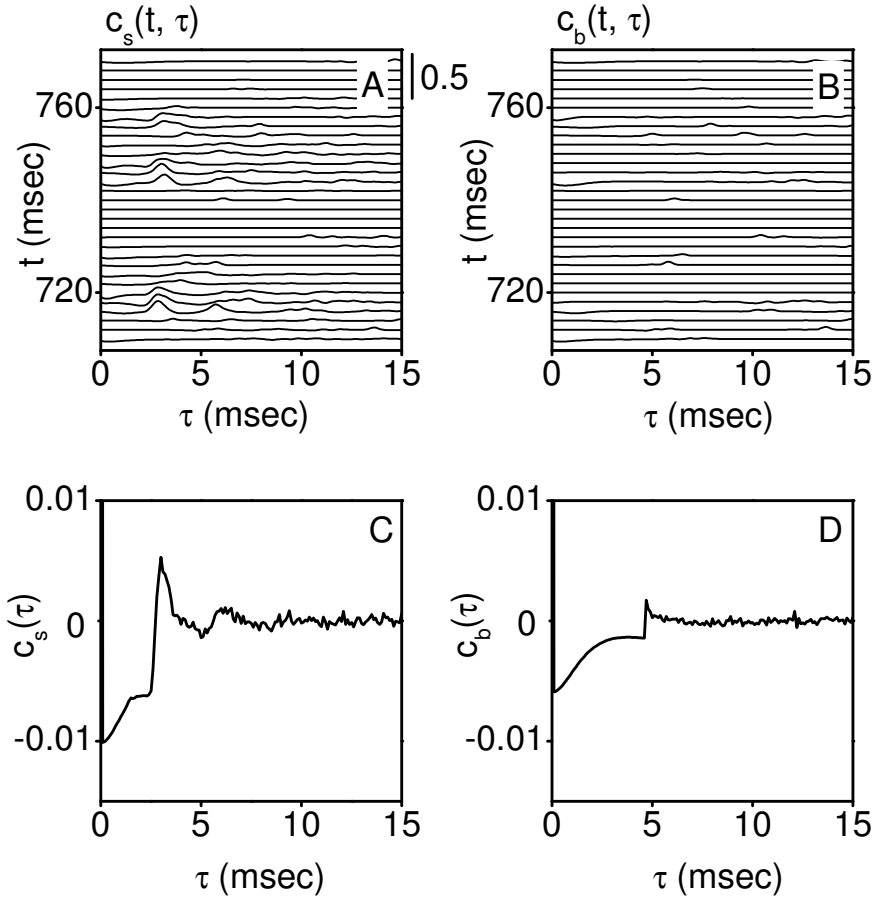}
\caption{\label{correl} Pearson correlation coefficient for a sample
cell. {\em A:} Coefficient $c_s(t, \tau)$ between spikes generated
at times $t$ and $t + \tau$. The scale (also valid for {\em B}) is
given in the upper-right corner. {\em B:} Coefficient $c_b(t, \tau)$
between bursts generated at times $t$ and $t + \tau$. {\em C} and
{\em D}: Coefficients $c_s(\tau)$ and $c_b(\tau)$ between spikes and
bursts, respectively. In {\em A} and {\em C}, a pronounced peak is
seen for $c_s$ at around $\tau = 3$ msec. In {\em C}, there is also
an initial negative plateau. These structures are markedly reduced
in $c_b$ ({\em B} and {\em D}), underscoring that generic spikes are
more correlated than classified bursts.}
\end{figure}

We represent the spike train as a sequence of non-negative integer
numbers $n$, each number indicating the intra-burst spike count of
the burst whose first spike falls in a small time window $[t, t +
\delta t]$ (see Fig.~\ref{fcodes}{\em B}, for an example). This
representation should be compared to the more typical binary
representation (Fig.~\ref{fcodes}{\em A}), where each digit in the
sequence indicates the presence or absence of a spike in the
relevant time bin. As shown in some of the examples of
Fig.~\ref{f0}, the binary representation often contains strong
temporal correlations. The very definition of an $n$-burst aims at
bundling highly correlated spikes into a single burst event. Hence,
the representation in terms of bursts necessarily reduces the
statistical dependence between different time bins, as seen in
Fig.~\ref{correl}. In panel {\em A} we show the Pearson correlation
coefficient $c_s(t, \tau)$ between spikes at times $t$ and $t +
\tau$ (see Methods), in an example cell. For comparison, panel {\em
B} exhibits the correlation coefficient $c_b(t, \tau)$ between {\em
bursts} at times $t$ and $t + \tau$ (see Methods), of the same spike
train. For small $\tau$ values, the plot in {\em A} shows a number
of peaks, that are absent in {\em B}. For the cell shown in
Fig.~\ref{correl}, the mean value of $c_s^2(t, \tau)$ averaged over
all $t \in [200 \ {\rm msec}, 990 \ {\rm msec}]$ and $\tau \in [0,
10 \ {\rm msec}]$ is 2.94 times larger than the corresponding mean
of $c_b^2(t, \tau)$. The population average of this ratio on all
bursting sessions is 2.89 (SD 1.49).


Figure~\ref{correl}{\em C} depicts the Pearson correlation
coefficient $c_s(\tau)$, averaged both over all trials and all times
$t$ (see Methods). For comparison, the Pearson correlation
coefficient $c_b(\tau)$ obtained with an $n$-burst representation of
the spike train is shown in {\em D}. The most prominent peak of
$c_s$ appears markedly diminished in $c_b$. This reduction
demonstrates that bursts are more independent from each other than
individual spikes.

Given the additive properties of information (Cover and Thomas,
1991), if in one particular case, a collection of events can be
shown to contain independent elements only, then the information
transmitted by the collection is the sum of the information
transmitted by the individual events. Figure~\ref{correl} shows that
the correlations between bursts are not strictly zero. Yet, if they
can be assumed to be negligible, and if there are no higher order
correlations, then the mutual information transmitted by the train
of bursts can be easily calculated from the information in small
time bins (see Methods, and Brenner et al., 2000).

\begin{figure}[htdf!]
\includegraphics{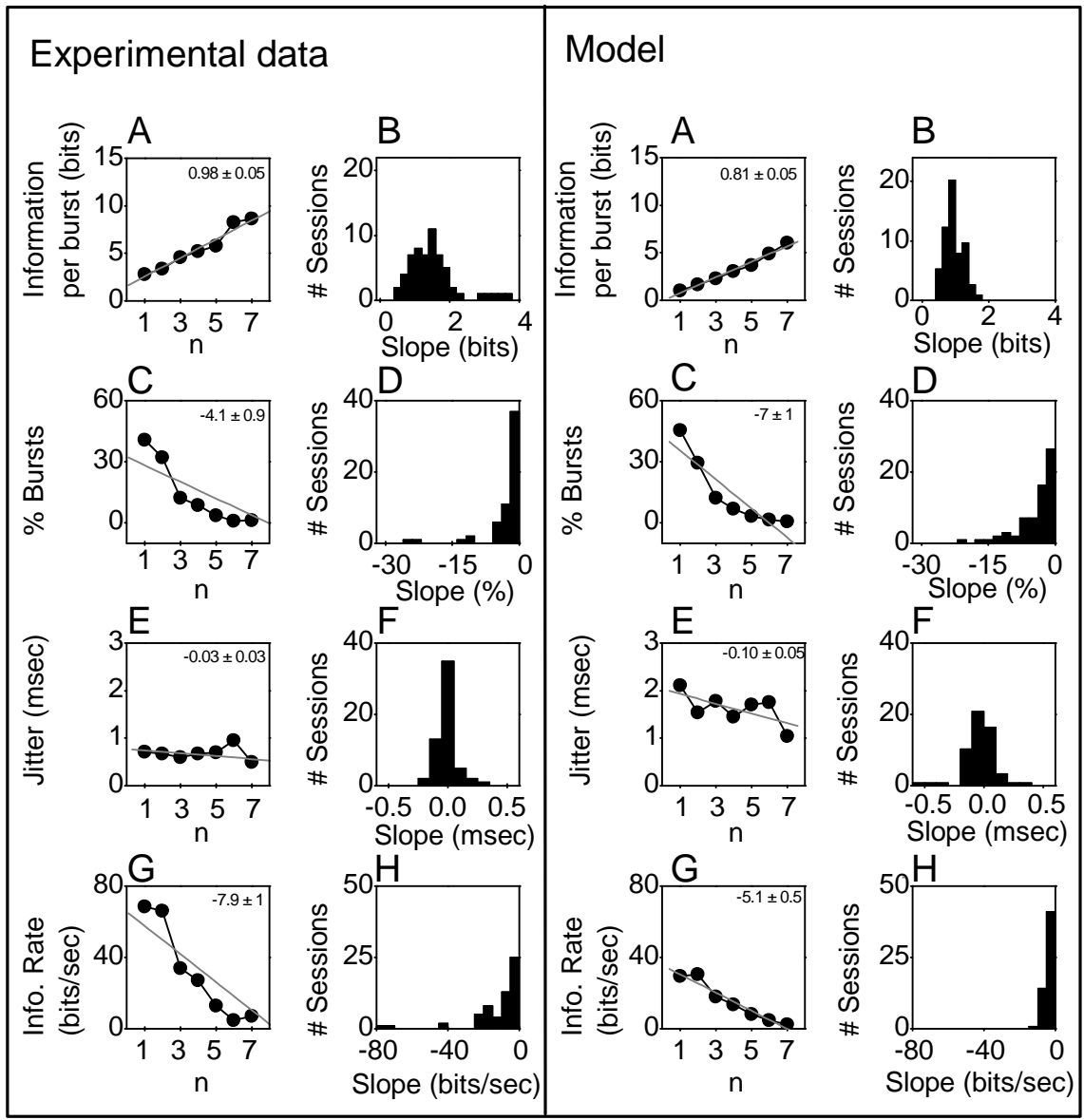}
\caption{\label{ffractions} Information transmitted in burst firing.
{\em Left half of the figure:} Experimental data. {\em Right half of
the figure:} Threshold linear model with refractoriness. {\em Left
column:} Data from the example cell of Fig.~\ref{f2}, with best
linear fits. Their slopes are given in the upper-right corner, with
their errors. {\em Right column:} Population data showing the
distribution of slopes of the linear fits for the quantities of the
left column. {\em A}: Average information transmitted by each
$n$-burst. The information transmitted per burst increases
monotonously with $n$. {\em B}: For all cells in the population, the
information per burst increases with $n$. {\em C}: Number of
occurrences of each $n$-burst. The larger the intra-burst spike
count $n$, the more rarely it appears. {\em D:} For all cells in the
population, low-$n$ bursts appear more frequently than high-$n$
bursts. {\em E}: Mean amount of jitter of the first spike of each
$n$-burst. {\em F:} The population data demonstrate that for some
cells, the amount of jitter is a slowly increasing function of $n$,
whereas for other cells, it is decreasing. {\em G}: Rate of
transmitted information for all $n$-bursts. Although isolated spikes
($n$ = 1) are the most frequent events (see {\em B}) a large
fraction of the transmitted information is carried by bursts. {\em
H:} For all cells in the population, the information rate decreases
with $n$. As shown by these data, the model captures the coding
trends of the investigated neurons. }
\end{figure}

In Fig.~\ref{ffractions}, the information transmitted by burst
firing is depicted for a sample cell, the left half of the figure
corresponding to the experimental data, the right half to the
threshold-linear model with refractoriness. Panel {\em A}
depicts the average information $I_n^{(1)}$ provided by each
$n$-burst. The higher the intra-burst spike count $n$, the more
informative the event is. To evaluate the significance of this
trend, we fitted the data with a straight line, and evaluated the
sign of the resulting slope, taking the estimated error bar of the
fit into account. In the upper right corner of panel {\em A}, the
value of the slope and its estimated error bar is indicated.
Since $n$ is dimensionless, slopes are also measured in bits. To
assess how often $I_n^{1}$ was increasing at the
population level, the analysis was repeated for all recorded
bursting cells. All sessions had significantly
positive slopes. Fig.~\ref{ffractions}{\em B} shows the
distribution of slopes throughout the population. The average slope
across the 59 bursting cells was 1.5 bits (SD 0.7 bits).


The information per burst $I_n^{(1)}$ is proportional to the
dissimilarity between the time-de\-pend\-ent probability density
$r_n(t)$ of an $n$-burst (see Methods) and a time-independent
distribution of the same mean rate $\bar{r}_n$. As such, it is large
whenever $r_n(t)$ is a highly uneven function of time, almost always
equal to zero, and only seldom exhibiting a sharp peak at a single,
or at most a few, particular values of $t$. A burst is therefore a
good candidate to transmit a large amount of in\-for\-ma\-tion per
event if it happens rarely (in each single trial), reliably (in a
large fraction of the trials), and with high temporal accuracy.
Panels {\em C} and {\em E} of Fig.~\ref{ffractions} depict the
frequency of occurrence $\bar{r}_n / \sum_n \bar{r}_n$ and the
amount of jitter of different $n$-bursts, respectively. Panel {\em
C} shows that high-$n$ bursts occur seldom. This result was also
observed in all other recorded cells: the frequency of occurrence
always decreased significantly with $n$. The population data in {\em
D} had an average slope of -2.6, with SD of 0.7. In {\em E}, the
amount of jitter in the first spike of the burst is shown to be
fairly constant with $n$. At the population level, in 80\% of the
bursting sessions the amount of jitter was roughly independent from
$n$ (the best linear fit had a slope that was not significantly
different from zero). The remaining 20\% showed a mild dependence,
but with no uniform trend, as shown by the population data in {\em
F}. The mean slope was -0.03 msec (SD 0.2 msec). The combined effect
of an event probability that diminishes strongly with $n$ (panel
{\em C}) and a jitter that is fairly constant with $n$ (panel {\em
E}) results in an information per event $I_n^{(1)}$ that increases
with $n$ (panel {\em A}).

The mutual information rate $I'_n$ of all $n$-bursts is proportional
to the product of the rate of $n$-bursts $\bar{r}_n$ and the mean
information transmitted by each $n$-burst $I^{1}_n$ (see Methods).
$I'_n$ strongly decreases with $n$ (Fig.~\ref{ffractions}{\em G}).
Similar results were obtained in all other recorded sessions ({\em
H}), with an average slope of -11 bits/sec (SD 16 bits/sec). The
total information rate $I'$ transmitted by the cell in
Fig.~\ref{ffractions} is, under the independence assumption, the sum
of all the columns in panel {\em G}, i.e. 220 bits/sec. Although
isolated spikes are the events transmitting information at the
highest rate, the collection of all $n > 1$ bursts, taken together,
provide no less than 69\% of the total information. The population
average of this fraction among all bursting cells was 47\%. Bursts,
therefore, constitute an important part of the neural code employed
by grasshopper auditory receptors.

The right half of Fig.~\ref{ffractions} shows the results obtained
for threshold linear model neurons with added refractoriness. For
each recorded cell a simulation was carried out, with the same
threshold, refractory period, and filter characteristics as the real
neuron. A comparison between the left and right panels of
Fig.~\ref{ffractions} reveals that the model reproduces the general
trends observed in the experimental data, both at the single-cell
and population level. Note that the model has no free fit
parameters.

The procedure introduced here allows one to calculate mutual
information rates between time-dependent stimuli and burst responses
in a straightforward fashion. However, apart from assuming
independence, the method contains one additional assumption. We have
grouped all bursts with $n$ spikes into one single type of event,
even if among those $n$-bursts there might be subtle differences in
the size of the ISIs. The first peak in the correlation function has
a certain width, so not all the spike doublets classified as a
2-burst are separated by exactly the same interval (see
Fig.~\ref{fcodes} for an example), and the same holds for all $n>1$.
If those differences were systematic, they could transmit additional
information about the stimulus. This type of information would be
lost through our procedure. We have, however, verified that
subsequent spikes inside a burst have larger amounts of jitter than
the first spike (data not shown). This suggests that the fine
temporal resolution in the spiking times of the subsequent spikes is
not crucial to information transmission.

In order to assess whether this is actually the case, we have
compared the information rates obtained with our procedure with
those resulting from the so-called {\em direct method} (Strong et
al., 1998). In this method, the spike train is segmented into binary
strings where the presence of a spike in a given time bin is
indicated by a 1, and silence is denoted by 0. A word is then
defined as a finite sequence of binary digits. The direct method
estimates the mutual information between stimuli and responses from
the probability distributions of all words of the spike train, in
the limit of large word lengths. This method has the advantage of
making no {\em a priori} assumptions about the neural code. The
drawback is that the size of the response space grows exponentially
with the length of the coding words. Due to sampling problems, in
our case it was therefore not possible to extend the maximal word
length beyond 3.2 msec (this includes no more than 2-bursts), with a
temporal precision equal to 0.4 msec. The sampling bias was
corrected using the NSB approach (Nemenman et al., 2004). The
information measures obtained by our method and by the direct method
were highly correlated ($R = 0.95$, using all sessions). The
population average obtained with the direct method is 222 $\pm$ 69
bits/sec. With our method, instead, this average was 191 $\pm$ 72
bits/sec. In all cases but one, the information obtained with the
direct method was higher than the one obtained with our method, the
average difference being 31 $\pm$ 16 bits/sec. It is still not clear
whether the remaining discrepancies are due to the cogency of the
assumptions raised by our method, or due to the limited word length
used in the direct method. If the direct method can be taken as a
reliable estimation, then by ignoring (a) the internal temporal
structure inside bursts and (b) the temporal correlations between
bursts, we are losing 14\% of the information. We emphasize,
however, that in contrast to the direct method, our procedure to
calculate information rates allows one to discriminate which
$n$-bursts are the most informative ones, and thereby, to gain a
better insight into the neural code.

\subsection{Qualitative description of the information transmitted by
bursts}

\label{qualitative}

The previous section shows that when the stimulus statistics is
varied, the probability of generating bursts of $n$ spikes varies
concurrently. To quantify the relevance of $n$-bursts for neural
coding, the mutual information rate associated with burst spiking
was calculated. Since bursts transmit information about the
stimulus, it should be possible to associate different stimuli with
different $n$ values. We now analyze this correspondence in detail.

There are two quantities of interest (Rieke et al., 1997). The first
one is the probability $P[n | s(\tau)]$ of finding an $n$-burst in
response to the stimulus $s(\tau)$. This quantity constitutes a
natural target in experimental studies that systematically explore a
given stimulus space. The second quantity is the probability
$P[s(\tau) | n]$ that a stimulus $s(\tau)$ was presented, given that
the cell generated an $n$-burst. This quantity is relevant for
reading out a neural code based on intra-burst spike numbers.

\begin{figure}[htdf!]
\includegraphics[scale = 0.9]{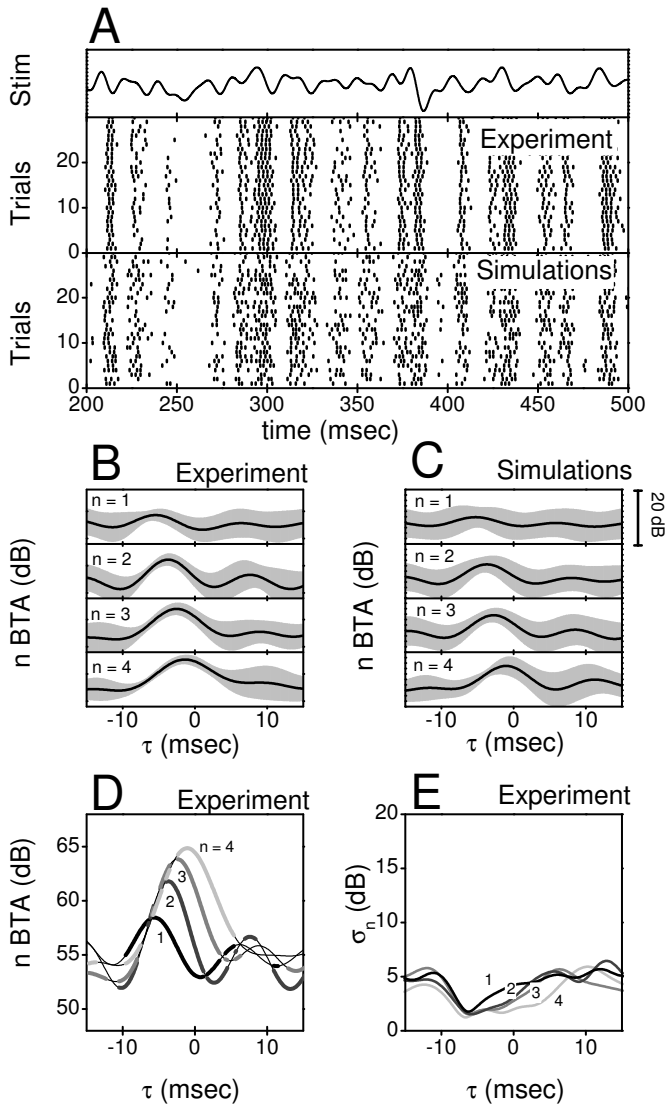}
\caption{\label{f5} Definition and characteristics of the $n$-burst
triggered averages ($n$BTA). {\em A}: Acoustic stimulus (top) and
the first 30 (out of 100) trials of the recorded (middle) and
simulated (bottom) neural responses. The AM signal had a standard
deviation of 6 dB and a cutoff frequency of 100 Hz. {\em B} and {\em
C}: All stimulus segments generating bursts of a given $n$ were
collected together and aligned with respect to the time of burst
initiation to obtain the $n$BTAs, shown for real ({\em B}) and
simulated ({\em C}) data. Grey areas represent the SD. {\em D}:
$n$BTAs as a function of time, for four different values of $n$.
Thick lines mark the segments where each $n$BTA is significantly
different from the other three, as assessed with a Student's t-test
($p < 0.01$). {\em E}: The standard deviation of each $n$BTA as a
function of time (see Eq.~\ref{snbta}). Approximately 7 msec before
the first spike of a burst is recorded, the standard deviation shows
a minimum, implying that at this moment the different stimuli
preceding an $n$-burst are most similar. This time lag was similar
for all $n$.}
\end{figure}

We begin by characterizing $P[s(\tau) | n]$. As an example,
Fig.~\ref{f5}{\em A} depicts 300 msec of an acoustic stimulus (upper
panel) and the corresponding neural (middle) and simulated (lower
panel) responses. The simulated threshold-linear neurons are clearly
less precise than the real receptor cells (see also
Fig.~\ref{ffractions}{\em E}). We then collected all stimulus
segments inducing burst generation, and aligned them such that burst
initiation was at $t = 0$. The $n${\em -burst triggered average}
nBTA is defined as the mean value of the aligned segments. In
Fig.~\ref{f5} {\em B} and {\em C}, $n$BTAs($t$) are depicted for the
experimental and simulated data, respectively. The grey areas
represent the SD of the average. Height and width of the $n$-BTA
increase with $n$. To determine whether this trend is significant,
the collections of stimulus segments corresponding to different $n$
values were compared with a two-way ANOVA test (see Methods). All
recorded and simulated bursting cells exhibited significantly
different $n$-BTAs, for $n$ ranging between 1 and 4. We therefore
determined the time intervals in which the different $n$BTAs
differed significantly from one another. For each point in time a
t-test was performed, assessing whether a given $n$BTA$(t)$ was
different from the $n'$BTA$(t)$ corresponding to other $n' \ne n$.
The result is shown in Fig.~\ref{f5}{\em D}. For those times $t$
where significant differences are found, the $n$BTA is represented
with a thick line. Most of the central peak in each $n$BTA is
significantly different from the other three curves. Notice that
both the height and the width of the most pronounced peak in the
$n$BTA increase systematically with $n$. Moreover, the mean delay
between stimulus upstroke and burst generation decreases
systematically with $n$. This implies that stimulus deflections that
are either high or wide tend to produce prompt responses, with
high-$n$ bursts. In what follows, the delay $\tau_n$ between the
maximum in each $n$BTA and the generation of an $n$- burst is called
burst latency.


The standard deviation $\sigma_n(\tau)$ of all stimuli generating
$n$-bursts provides a measure of the dissimilarity between the
stimulus segments. If there is a particular $\tau$ for which
$\sigma_n(\tau)$ becomes markedly small, then, for that time $\tau$,
the stimuli preceding an $n$-burst are noticeably similar to each
other. In Fig.~\ref{f5}{\em E}, $\sigma_n(\tau)$ is depicted. There
is a clear minimum approximately 7 msec before burst generation,
coinciding with the sharp upstroke in the $n$BTA. This delay
includes sound propagation ($\approx 1$ msec) and axonal delays
($\approx$ 2 msec). Notice that the position of this minimum remains
roughly unchanged, as $n$ is varied. Its standard deviation for
different $n$ values is 0.33\ msec, for this cell. The constancy of
the location of the minima also holds at the population level. The
mean standard deviation of the position of the minima of
$\sigma_n(\tau)$ was roughly 0.05 times the inverse cutoff
frequency. Its average among all bursting cells is 0.4 msec (SD 0.7
msec) considering $1 \le n \le 4$.

For 98\% of the bursting cells and for all $n$ values,
$\sigma_n(\tau)$ was small\-er than the standard deviation
$\sigma(\tau)$ of the stimuli preceding all spikes (prior to any
classification). The population average of the ratio of the minimum
value of $\sigma(\tau)$ to the $n$-average of the minimum values of
$\sigma(\tau)$ was 1.62 (SD 0.56). The set of stimuli preceding all
spikes thus constitutes a more heterogeneous collection than the set
of stimuli preceding an $n$-burst. This is not surprising. If, say,
a burst of 3 spikes is systematically generated after one particular
stimulus feature, the spike-triggered average includes three
time-shifted copies of the relevant feature. This threefold
collection of stimuli has a larger standard deviation than the set
of stimuli preceding a 3-burst. In a related study (Gollisch, 2006),
spike-time jitter was shown to broaden the STA. Our data demonstrate
that burst firing, although not necessarily accompanied by jitter,
gives rise to a similar effect. Therefore, whenever the tendency to
fire bursts is high, the collection of stimuli preceding {\em spike}
generation may show a large variance, rendering the interpretation
of the STA of little use. In these cases, the burst-triggered
average may provide additional insight.

Not all bursting cells display $n$BTAs as those shown in
Fig.~\ref{f5}{\em D}. In some cases, for example, the central peak
of the 4BTA is slightly lower than that of the 3BTA, though markedly
wider. These differences reflect individual properties of different
neurons. However, out of the 58 sessions where bursting cells were
found, 50 exhibit $n$BTAs whose central peaks were significantly
different from one another---except, of course, at those points
where the curves cross. The remaining 8 sessions corresponded to
cases where bursts appeared only seldom, thereby contributing with a
number of samples that was too small to assess significant
differences.

\begin{figure}[htdf!]
\includegraphics{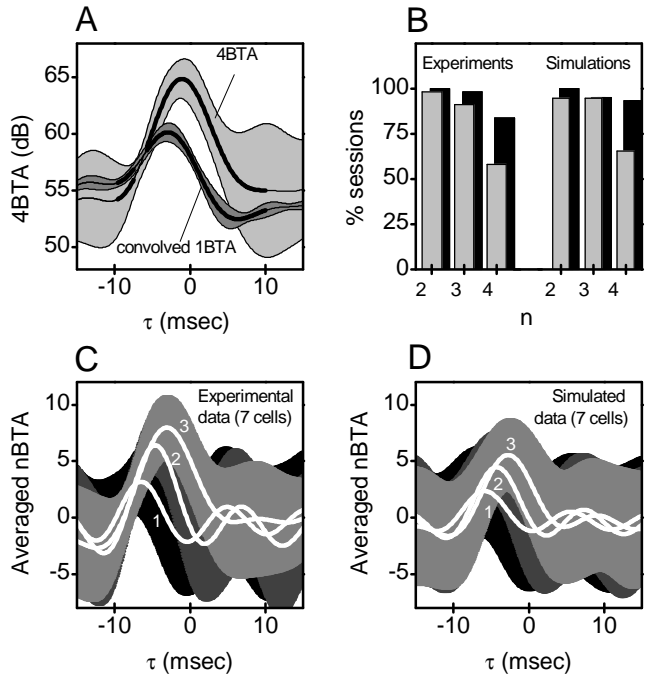}
\caption{\label{autoval} Analysis of the $n$BTAs. {\em A}:
Comparison between the 4BTA obtained for the cell depicted in
Fig.~\ref{f5}{\em D} and the function that results from convolving
four 1BTAs interspaced by the ISIs found in the real data. Thick
lines denote segments that differ significantly between the two
cases. Shaded areas represent the standard deviation of the averaged
data. The estimated error of the 4BTA and the convolved 1BTA is
approximately 10 times smaller than the SD of the averaged data.
{\em B:} Black bars: Percentage of cells for which the real $n$BTA
differs from the convolved 1BTA anywhere inside the time interval
between the two minima at each side of the maximum in the $n$BTA, as
assessed by a two-way ANOVA test. Grey bars: percentage of cells
where a significant difference was found in at least 70\% of the
tested time interval, as assessed by a point-by-point Student
t-test. {\em C}: Population average of the 1BTA, 2BTA, and 3BTA for
the 7 cells driven with an AM signal with 100 Hz cutoff frequency
and 6 dB standard deviation. Shaded areas represent the SD of the
average. Black: SD of the 1BTA. Grey: SD of the 2BTA. Light grey: SD
of the 3BTA. {\em D:} Same as {\em C}, but obtained from simulated
threshold-linear cells with refractory period. }
\end{figure}

A burst is a sequence of shortly interleaved spikes. Could the
$n$BTAs obtained for high $n$-values shown in Fig.~\ref{f5}{\em D}
have been obtained by combining a sequence of $n$ interspaced 1BTAs,
or even spike-triggered averages (STAs)? To answer this question, in
Fig.~\ref{autoval}{\em A} we compare the same 4BTA depicted in
Fig.~\ref{f5}{\em D} with a curve obtained by combining four 1BTAs
interspaced with the ISIs found in the real data. The shaded areas
represent the SD of the averaged data. We see that the two curves
are clearly different from each other, the real 4BTA being markedly
higher and wider than the combined 1BTAs. This implies that the
stimulus deflections triggering bursts of $n$ = 4 are significantly
higher than those required to generate four spikes of $n = 1$.


To test other cells in the population for the same effect, for each
$n$ we determined the fraction of sessions for which the $n$BTA
differed significantly from the combined 1BTAs (or STAs) in an
interval extending between the two minima at each side of the
central maximum of the $n$BTA. This comparison was done by means of
a two-way ANOVA (see methods). Black bars depict the fraction of
cells where a significant difference was found. Among the cells that
exhibited significant differences, we tested whether the difference
could be observed in a substantial fraction of the tested interval.
To that end, we carried out a Student t-test for each time point
within the time interval between the two minima at each side of the
maximum of the $n$BTA (see Methods). We counted the cells showing
significant differences in more than the 70\% of the tested
interval. The results are depicted in grey bars in
Fig.~\ref{autoval}{\em B}. A large fraction of the cells show a
significant difference, for both real and simulated data. Hence,
also at the population level, the $n$BTAs differ significantly from
the convolved 1BTAs. As $n$ increases, the number of sessions with
significant differences diminishes. This is a consequence of the
fact that for larger $n$, there are fewer $n$-bursts, and therefore,
the error bar of the estimation of the $n$BTA increases. As an
additional check, we repeated the analysis by convolving $n$ shifted
copies of the STA, instead of the 1BTA, obtaining similar results.

Finally, we checked that for the same stimulus, the $n$BTAs of
different cells showed a similar trend, as $n$ varied. The
population average was taken after subtracting the mean stimulus to
each $n$BTA because different receptors were recorded with different
mean stimuli (see Methods). In addition, since different cells
showed different latencies $\tau_n$, all stimulus segments were
shifted by $\tau_n$ before averaging, and then shifted back
afterwards. Fig.~\ref{autoval}{\em C} demonstrates that also at the
population level, high-$n$ bursts are associated with either higher
or wider stimulus deflections (or both). The large error bars
indicate that there is no absolute value of a stimulus fluctuation
that uniquely triggers bursts of a given $n$ value, throughout the
population. The qualitative behavior is also reproduced by the
threshold-linear model with refractory Period
(Fig.~\ref{autoval}{\em D}). We conclude that both in real and
modeled data, high-$n$ bursts are associated with high or wide
stimulus deflections.

\begin{figure}[htdf!]
\includegraphics[keepaspectratio=true, clip = true,
scale = 1, angle = 0]{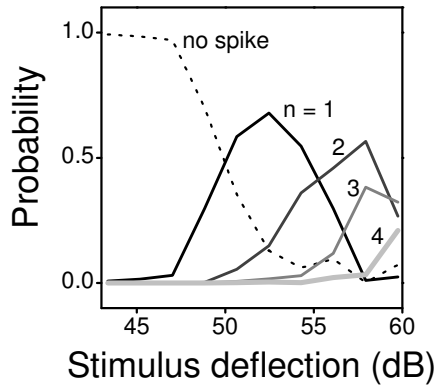} \caption{\label{probability}
Probability of generating no response (dotted line), an isolated
spike (black, solid line), a spike doublet (dark grey) or triplet
(grey), or a burst with 4 spikes (light grey line) as a function of
the height of the stimulus deflection, for an example cell. The
uncertainties of these probabilities have been estimated (Samengo,
2002), and the absolute error was always below 0.025. The
segregation between the lines indicates that the number of spikes in
a burst contains information about the height of the stimulus upward
excursion preceding the burst.}
\end{figure}

Let us turn to the analysis of $P[n | s(\tau)]$ and describe how
this quantity varies with the height of the deflections in
$s(\tau)$. The shape of the $n$BTAs (Fig.~\ref{f5}{\em D})
demonstrates that the average stimulus preceding an $n$-burst always
contained a prominent up-and-down excursion, whose maximum was
located some $\tau_n$ milliseconds before burst initiation. This
indicates that there is an association between upward stimulus
excursions and burst generation. Can we assert that the probability
of generating a burst of $n$ spikes at a given time depends on the
size of the upward stimulus excursion? In order to explore this
question, we estimated the probabilities $P(n | h)$ of obtaining
bursts of $n$ spikes following an upward stimulus excursion of
height $h$ (see Methods). For an example cell,
Fig.~\ref{probability} shows a marked segregation between the
responses elicited by deflections of different heights. Whereas
fairly low excursions produce either no response (dotted line) or an
isolated spike (black, solid line), large deflections are associated
with doublets (dark grey), triplets (grey), or bursts with 4 spikes
(light grey line). All cells showing a bursting behavior exhibited
this phenomenon.


\section{Discussion}

\noindent The role of burst firing for neural coding has been
studied extensively in systems where individual neurons have an
intrinsic tendency to burst. Typical examples are electrosensory
neurons of electric fish (Metzner et al., 1998; Oswald et al., 2004)
and thalamic relay cells in the visual systems of cat (Ailitto et
al., 2005; Denning and Reinagel, 2005; Lesica et al., 2006) and
mouse (Grubb and Thompson, 2005). For downstream neurons, however,
it is irrelevant {\em how} bursts are generated. All that matters is
their representational properties, i.e., their structure and coding
capability. Therefore, we have focused on the coding properties of
cells that lack intrinsic burst mechanisms. In particular, we wanted
to know how much sensory information is transmitted and which
symbols in the neural code are associated with each stimulus
feature. To that end, we analyzed the activity of grasshopper
auditory receptor neurons and simulated neurons, both lacking
intrinsic bursting mechanisms. We first introduced a criterion that
allowed us to determine the cases where a neural response could be
considered as a sequence of bursts. Next, we explored a code based
on the intra-burst spike count $n$. We estimated the information
transmitted by this code, and characterized the correspondence
between specific stimulus features and specific $n$ values. We
observed that long bursts are associated with particularly high or
long stimulus excursions, and that this effect could not be
reproduced by concatenating the stimuli generating short bursts. In
the following subsections, we discuss our results in the context of
previous studies.

\noindent{\bf Burst identification benefits from considering neural
response statistics}

\noindent
In previous analyses, burst identification typically relied
on strict boundaries on the ISIs (see, for example, Alitto et al.,
2005; Denning and Reinagel, 2005; Lesica and Stanley, 2004; Oswald
et al., 2004). This is appropriate for cells that have intrinsic
burst mechanisms with fairly rigid time constants. However, neurons
that do not burst intrinsically exhibit intra-burst ISIs of variable
duration, depending on the temporal properties of the stimulus as
shown by a comparison of the peak widths in figures 3B and 3G.
Hence, in this work the criterion used to determine whether two
consecutive spikes were or were not part of a burst was uniquely
tailored for each session. Note that if a cell is classified as
non-bursting, this does not imply that it does not generate
bursts at all, but rather, that the intra-burst ISIs (if present)
cannot be cleanly separated from the inter-burst ISIs. In these
cases it is not possible to interpret the neural code in terms of
distinct words formed by closely spaced spikes.

\noindent{\bf Burst coding does not require intrinsic burst
dynamics}

\noindent Not all cells investigated in this study were bursters:
Some cells bursted in response to some stimuli, and responded
tonically to other stimuli. Indeed, grasshopper receptors do not
burst when driven with constant or step stimuli (Gollisch et al.,
2002; Gollisch and Herz, 2004). In other studies, the time-scales
of stimuli eliciting bursts have often been related to the
particular ionic currents involved in burst generation (Alitto et
al., 2005; Denning and Reinagel, 2005; Dorion et al., 2007; Krahe
and Gabbiani, 2004; Lesica et al., 2006). Oswald et al. (2004)
also presented a mathematical model in which bursts were only able
to support efficient feature detection when a specific active
dendritic backpropagation was present. Our results, however,
demonstrate that burst-coding does not require complex intrinsic
neural dynamics, as shown by our minimal computational model (see
Keat et al., 2001, for another example). Although simulated
neurons were in general less precise than real neurons, they
showed similar correlation functions (Fig.~\ref{f2}), and coding
properties (Figs.~\ref{ffractions} and \ref{f5}). These findings
underscore that the tendency to burst does not need to be an
intrinsic propensity of the cell per se, but may arise as a
consequence of how its cellular properties interact with the
temporal characteristics of the external stimulus. Our system,
therefore, is an example of {\em stimulus-induced bursting} as
previously reported by Neiman et al. (2007).

\noindent{\bf Comparison with other neural codes}

\noindent We have assumed that the relevant code symbols are the
time at which a burst is initiated, and the intra-burst spike count
$n$. There are, however, other burst-based neural codes that have
been explored previously. For example, Kepecs et al. (2002,
unpublished) reported that the relevant information can be encoded
in the total duration of a burst. In the cells of our study, $n$ was
proportional to burst duration (data not shown). This implies that
for those neurons a code based on the intra-burst spike count $n$ is
equivalent a burst-duration code. On the other hand, in
electrosensory neurons of electric fish, ISIs in bursts with two
spikes depend on the amplitude of electric-field upstrokes was
encoded in the duration of ISIs of bursts of $n = 2$ (Oswald et al.,
2007). Grasshopper auditory receptors, however, have rather narrow
ISI range of at most 3 msec, to be compared with the typical range
of 8 msec in electric fish. We have therefore not explored a code
utilizing the intra-burst ISI length.

Previous studies have also reported $n$-based neural codes in
different sensory modalities. In visual cortex, for example, $n$
depends on stimulus orientation, as shown by De Busk et al. (1997),
Martinez Conde et al. (2002) and others. In the vertebrate retina,
$n$ carries information about the stimulus history preceding burst
initiation (Berry et al., 1997). Experimental data from cat LGN
(Kepecs et al., 2001; Kepecs et al., unpublished) and computational
models (Kepecs et al., 2002) demonstrate that $n$ can encode the
slope of stimulus upstrokes.

We would like to emphasize that an $n$-burst code differs from a
firing-rate code. Within a firing-rate code, each point in time is
associated with a specific time-dependent firing rate. This rate may
be computed as an instantaneous firing rate from local ISIs, or by
convolving the spike train with a certain filter function. In either
case, the precise time course of the original spike train may be
fully recovered. This is not true for the $n$-burst code, where
information about the exact spike times within each burst is lost -
in essence, the code only looks at whether there is a spike within
the time interval defined through the correlation function, or not.
Thus, the $n$-burst code provides a highly reduced representation,
and not a full firing-rate code.

Our analysis shows, however, that in spite of this reduction the
$n$-burst code still contains a large fraction (approximately 85\%)
of the total transmitted information, as deduced from comparing our
results with the direct method. In addition, by parsing the
responses into code-words, the code is amenable for read-out. Our
results show significant differences between the stimuli encoded by
different $n$-values and reveal those stimuli explicitly.

\noindent {\bf Implications for the neural code}

\noindent We have also derived a procedure to calculate the mutual
information rate between stimuli and responses if different bursts
can be assumed to be independent from each other. This technique
should be extended with caution to other systems since the small
size of inter-burst correlations found in grasshopper auditory
receptors may not be shared by other sensory systems. In addition,
vanishing inter-burst correlations do not guarantee that the bursts
be independent. Higher-order correlations could still be present.
Our approximation assumes that those terms can be neglected when
computing information measures.

The consequence of assuming that different $n$-bursts are
independent from one another is that the total transmitted
information may be decomposed into the sum of the information
transmitted by each $n$-burst. This allows one to quantify which
$n$-values are most relevant. Our data show that $n$-bursts with $n
> 1$ can transmit at least the same amount of information as
isolated spikes ($n$ = 1).

To analyze the relation between particular $n$-values and the
stimuli represented by these bursts, we calculated burst-triggered
averages for each $n$. The set of stimuli preceding different $n$
values differed significantly from one another. Specifically, $n$
was shown to be reliably associated to the height of the stimulus
upstroke preceding burst generation. In some cells, a weak
dependence on the width of the amplitude deflection, its slope, and
its integral was observed, too (data not shown). However, at the
population level, the stimulus feature that most reliably co-varied
with $n$ was the maximal height of the AM signal.

The two aspects that seem to be most relevant for information
transmission, i.e. the time at which a burst is initiated and the
intra-burst spike count $n$, would also be good candidates to
represent what in the literature has been distinguished as the {\em
when} and the {\em what} in a stimulus (Berry et al., 1997; Borst
and Theunissen, 1999; Theunissen and Miller, 1995). In our data,
bursts containing different numbers of spikes are associated with
sound fluctuations of different heights and widths. The $n$-value
thus provides qualitative information about two key stimulus
aspects. In addition, the time at which a burst begins indicates
when the corresponding acoustic feature occurred. Notice that both
aspects are interwoven, because the response latency decreases with
increasing $n$. To decode the precise arrival time of an acoustic
signal, downstream neurons therefore also need to read out the
intra-burst spike count $n$. This provides additional independent
evidence for the usefulness of the $n$-burst code investigated in
this study.

\pagebreak

\section*{Acknowledgments}

\noindent We are grateful to Adam Kepecs for sharing his unpublished
manuscript {\em A burst-duration code in the thalamus} with us. We
thank Tim Gollisch and Germ\'an Mato for useful discussions.

\noindent The present address of A. Rokem is: Helen Wills
Neuroscience Institute, University of California, Berkeley, CA
94720, USA. The present address of A.V.M. Herz is: Biozentrum,
Ludwig-Maximilans-Universit\"at M\"unchen, 82152
Planegg-Martinsried, Germany.

\noindent This work was supported by the Alexander von Humboldt
Foundation, the Deutsche Forschungsgemeinschaft (SFB 618), the
Consejo de Investigaciones Cient\'{\i}ficas y T\'ecnicas, the German
Federal Ministry of Education and Research, the Israeli Ministry of
Science, the Minerva Foundation of the Max Planck Society, and the
Secretar\'{\i}a de Ciencia y Tecnolog\'{\i}a of Argentina.

\pagebreak

\section*{References}

\begin{description}


\item Ag\"uera y Arcas B, Fairhall AL, Bialek W (2003). Computation
in a Single Neuron: Hodgkin and Huxley Revisited. Neural Comput 15:
1715–-1749.

\item Alitto HJ,  Weyand TG, Usrey WM (2005). Distinct properties
of stimulus-evoked bursts in the lateral geniculate nucleus. J
Neurosci 25: 514--523.



\item Barlow RJ (1999). Statistics: a guide to the use of
statistical methods in the physical sciences. West Sussex, Wiley.

\item Benda J, Bethge M, Hennig R M, Pawelzik K, Herz AVM (2001).
Spike-frequency adaptation: phenomenological model and experimental
tests. Neurocomput 38--40: 105--110.

\item Benda J, Herz AVM (2003). A Universal Model for Spike-Frequency
Adaptation. Neural Comput 15: 2523-–2564.

\item Berry MJ, Warland DK, Meister M (1997). The structure and
precision of retinal spike trains. Proc Natl Acad Sci U S A 94:
5411–-5416.

\item Borst A, Haag J (2001). Effects of mean firing on neural
information rate. J Comput Neurosci 10: 213--221.

\item Borst A, Theunissen FE (1999). Information theory and
neural coding. Nat Neurosci 2: 947--957.

\item Brenner N, Strong SP, Koberle R, Bialek W, de Ruyter van
Steveninck RR (2000). Synergy in a Neural Code. Neural Comput 12: 1531--1552.


\item Chacron MJ, Longtin A, Maler L (2004). To burst or not to
burst? J Comput Neurosci 17: 127--136.

\item Cover T, Thomas J (1991). Elements of information theory.
New York: Wiley and sons.

\item DeBusk BC, DeBruyn EJ, Snider RK, Kabara JF, Bonds AB
(1997). Stimulus-Dependent Modulation of Spike Burst Length in Cat Striate Cortical Cells. J
Neurophysiol 78: 199–-213.

\item Denning KS, Reinagel P (2005). Visual Control of Burst
Priming in the Anesthetized Lateral Geniculate Nucleus. J Neurosci
25: 3531--3538.

\item Doiron B, Oswald AMM, Maler L (2007). Interval Coding. II.
Dendrite-Dependent Mechanisms. J Neurophysiol 97: 2744–-2757.

\item Eggermont JJ, Smith GM (1996). Burst-firing sharpens
frequency-tuning in primary auditory cortex. Neuroreport 7:
753--757.

\item Gollisch T (2006). Estimating Receptive Fields in the
Presence of Spike-Time Jitter. Network: Computat Neu Sys
17:103--129.

\item Gollisch T and Herz AVM (2005). Disentangling
Sub-Millisecond Processes within an Auditory Transduction Chain.
PLoS Biology 3(1):e8.

\item Gollisch T, Herz AVM (2004). Input-driven components of
spike-frequency adaptation can be unmasked in vivo. J. Neurosci.
24: 7435--7444.

\item Gollisch T, Sch\"utze H, Benda J, Herz AVM (2002). Energy
Integration Describes Sound-Intensity Coding in an Insect Auditory System. J Neurosci 22(23):
10434–-10448.

\item Gour\'evitch B, Eggermont JJ (2007). A nonparametric
approach for detection of bursts in spike trains. J Neurosci
Methods 160: 349–-358.

\item Grubb MS, Thompson ID (2005). Visual Response Properties of
Burst and Tonic Firing in the Mouse Dorsal Lateral Geniculate
Nucleus. J Neurophysiol 93: 3224--3247.

\item Hill K G (1983). The physiology of locust auditory receptors.
I: Discrete depolarizations of receptor cells. J Comp Physiol A 152:
475--482.

\item Izhikevich EM (2000). Neural Excitability, Spiking, and Bursting.
Int J Bif Chaos 10: 1171--1266.

\item Izhikevich EM and Hoppensteadt F.C. (2004). Classification
of Bursting Mappings. Int J Bif Chaos, 14: 3847--3854.


\item Keat J, Reingagel P, Reid RC, Meister M (2001). Predicting
Every Spike: A Model or the Responses of Visual Neurons. Neuron
30: 803–-817.

\item Kepecs A, Lisman J (2003). Information encoding and
computation with spikes and bursts. Network: Comput Neu Sys 14:
103--118.

\item Kepecs A, Lisman J (2004). How to read a burst duration
code. Neurocomput 58--60: 1--6.

\item Kepecs A, Sherman S, Lisman J (2001). Burst duration
coding in cat LGN. Soc Neurosci Abstr 31:13295.

\item Kepecs A, Wang XJ, Lisman J (2002). Bursting Neurons Signal Input
Slope. J Neurosci 22(20):9053--9062.

\item Koch C, Segev I (1998). Methods in Neural Modelling.
Cambridge Massachusetts, MIT Press.

\item Krahe R, Gabbiani F (2004). Burst firing in sensory
systems. Nat Rev Neurosci 5: 13--23.



\item Lesica NA, Stanley GB (2004). Encoding of Natural Scene
Movies by Tonic and Burst Spikes in the Lateral Geniculate Nucleus.
J Neurosci 24: 10731--10740.

\item Lesica NA, Weng C, Jin J, Yeh CI, Alonso JM, Stanley GB
(2006). Dynamic Encoding of Natural Luminance Sequences by LGN Bursts. PlosBiology 4 (7):e209.


\item Machens CK, Stemmler MB, Prinz P, Krahe R,
Ronacher B, and Herz AVM (2001). Representation of Acoustic
Communication Signals by Insect Auditory Receptor Neurons. J
Neurosci 21: 3215-–3227


\item Machens CK, Gollisch T, Kolesnikova O, Herz AVM (2005).
Testing the efficiency of sensory coding with optimal stimulus
ensembles. Neuron, 47: 447--456.


\item Martinez-Conde S, Macknik SL, Hubel DH (2002). The function
of bursts of spikes during visual fixation in the awake primate
lateral geniculate nucleus and primary visual cortex. Proc Nac Acad
Sci 99: 13920–-13925.

\item Metzner W, Koch C, Wessel R, Gabbiani F (1998). Feature
Extraction by Burst-Like Spike Patterns in Multiple Sensory Maps. J
Neurosci 18: 2283--2300.

\item Neiman AB, Yakusheva TA,  Russell DF (2007). Noise-Induced
Transition to Bursting in Responses of Paddlefish Electroreceptor
Afferents. J Neurophysiol 98: 2795--2806.

\item Nemenman I, Bialek W, de Ruyter van Steveninck RR (2004).
Entropy and information in neural spike trains: Progress on the
sampling problem. Physl Rev E, 69: 056111.

\item Oswald AMM, Chacron MJ, Dorion B, Bastian J, Maler L (2004).
Parallel Processing of Sensory Input by Bursts and Isolated Spikes.
J Neurosci 24: 4351--4362.

\item Oswald AMM, Doiron B, Maler L (2007). Interval Coding. I.
Burst Interspike Intervals as Indicators of Stimulus Intensity. J
Neurophysiol 97: 2731--2743.

\item Perkel DR, Gerstein GL, and Moore GP (1967).
Neuronal spike trains and stochastic point processes: I. The single
spike train. Biophysical J, 7, 391--418.

\item Reinagel P, Godwin D, Sherman SM, Koch C (1999). Encoding
of Visual Information by LGN Bursts. J Neurophysiol 81: 2558--2569.

\item Reich DS, Mechler F, Purpura KP, Victor JD (2000).
Interspike Intervals, Receptive Fields, and Information Encoding in
Primary Visual Cortex. J Neurosci 20(5): 1964–-1974.

\item Rieke F, Warland D, de Ruyter van Steveninck R, Bialek W
(1997). Spikes: Exploring the Neural Code. Cambridge, MIT Press.

\item Rokem A, Watzl S, Gollisch T, Stemmler M, Herz AVM, Samengo
I (2006). Spike-Timing Precision Underlies the Coding Efficiency of Auditory Receptor Neurons. J
Neurophysiol 95: 2541--2552.

\item R\"omer H. (1976). Die informationsverarbeitung typmanaler
Rezeptorelemente von Locusta migratoria (Acrididae,. Orthoptera). J
Comp Physiol A 109: 101--122.

\item Ronacher B, R\"omer H (1985). Spike synchronization of
tympanic receptor fibres in a grasshopper (Chorthippus biguttulus
L., Acrididae). J Comp Physiol A 157: 631-–642.

\item Samengo I (2002). Estimating probabilities from experimental
frequencies. Phys Rev E, 65: 046124.

\item Schaette R, Gollisch T, Herz AVM (2005). Spike-train variability
of auditory neurons in vivo: dynamic responses follow predictions
from constant stimuli. J Neurophysiol 93: 3270--81.


\item Sherman SM (2001). Tonic and burst firing: dual modes of
thalamocortical relay. Trends Neurosci 24: 122--126.

\item Sippel M, Breckow J (1983). Non-linear analysis of the
transmission of signals in the auditory system of the migratory
locust Locusta migratoria. Biol Cybern 46: 197-–205.

\item Strong SP, Koberle R, de Ruyter van Steveninck RR, Bialek W
(1998). Entropy and Information in Neural Spike Trains. Phys Rev Lett 80: 197--200.

\item Theunissen F Miller JP 1995. Temporal encoding in nervous
systems: a rigorous definition. J Comput Neurosci 2: 149--162.

\item von Helversen D, von Helversen O (1994). Forces driving
coevolution of song and song recognition in grasshoppers.
Fortschritte der Zoologie 39: 253--284.

\item Wang XL, Rinzel J (1995). Oscillatory and bursting
properties of neurons. In: MA Arbib, ec. The Handbook of Brain
Theory and Neural Networks, Cambridge Massachusetts, The MIT Press,
pp. 686--691.

\end{description}
\normalsize

\end{document}